# A Framework for FAIR and CLEAR Ecological Data and Knowledge: Semantic Units for Synthesis and Causal Modelling

Vogt, Lars[1]; König-Ries, Birgitta[2]; Alamenciak, Tim[3]; Brian, Joshua I.[4]; Arnillas, Carlos Alberto[5]; Korell, Lotte [6,7], Frühstückl, Robert[8]; Heger, Tina[9,10,11]

[1] *TIB Leibniz Information Centre for Science and Technology, Welfengarten 1B, 30167 Hanover, Germany,*  orcid.org/0000-0002-8280-0487

[2] *Institute for Computer Science, Friedrich Schiller University Jena, Fürstengraben 1, 07743 Jena, Germany,*  orcid.org/0000-0002-2382-9722

[3] *Department of Biology, Carleton University, 1125 Colonel By Drive, Ottawa, Ontario, Canada,*  orcid.org/0000-0002-1296-2528

[4] *Department of Geography, King's College London, London WC2B 4BG, United Kingdom,*  https://orcid.org/0000-0001-9338-4151

[5] *Department of Physical and Environmental Science, University of Toronto at Scarborough M1C 1A4, Ontario, Canada,* https://orcid.org/0000-0001-9338-4151

[6] *Helmholtz Centre for Environmental Research (UFZ), Theodor-Lieser-Str. 4, 06120, Halle (Saale)*

[7] *German Centre for Integrative Biodiversity Research (iDiv) Halle-Jena-Leipzig, Puschstr. 4, 04103 Leipzig, Germany*

[8] *Institute of Philosophy and Scientific Method, Johannes Kepler University Linz, Altenbergerstraße 69, 4040 Linz, Austria,*  https://orcid.org/0009-0005-3773-1039

[9] *Leibniz Institute of Freshwater Ecology and Inland Fisheries (IGB), Müggelseedamm 310, 12587 Berlin, Germany*  https://orcid.org/0000-0002-5522-5632

[10] *Freie Universität Berlin, Institute of Biology, Königin-Luise-Str. 1–3, 14195 Berlin, Germany*

[11] *Technical University of Munich, Germany; TUM School of Life Sciences, Restoration Ecology, Emil-Ramann-Str. 6, 85354 Freising, Germany*

Correspondence to: lars.m.vogt@googlemail.com

# Abstract

Ecological research increasingly relies on integrating heterogeneous datasets and knowledge to explain and predict complex phenomena. Yet, differences in data types, terminology, and documentation often hinder interoperability, reuse, and causal understanding. We present the Semantic Units Framework, a novel, domain-agnostic semantic modelling approach applied here to ecological data and knowledge in compliance with the FAIR (**F**indable, **A**ccessible, **I**nteroperable, **R**eusable) and CLEAR (**C**ognitively interoperable, semantically **L**inked, contextually **E**xplorable, easily **A**ccessible, human-**R**eadable and -interpretable) Principles.

The framework models data and knowledge as modular, logic-aware semantic units: single propositions (statement units) or coherent groups of propositions (compound units). Statement units can model measurements, observations, or universal relationships, including causal ones, and link to methods and evidence. Compound units group related statement units into reusable, semantically coherent knowledge objects. Implemented using RDF, OWL, and knowledge graphs, semantic units can be serialized as FAIR Digital Objects with persistent identifiers, provenance, and semantic interoperability.

We show how universal statement units build ecological causal networks, which can be composed into causal maps and perspective-specific subnetworks. These support causal reasoning, confounder detection (back-door), effect identification with unobserved confounders (front-door), application of do-calculus, and alignment with Bayesian networks, structural equation models, and structural causal models.

By linking fine-grained empirical data to high-level causal reasoning, the Semantic Units Framework provides a foundation for ecological knowledge synthesis, evidence annotation, cross-domain integration, reproducible workflows, and AI-ready ecological research.

# Introduction

As researchers and scientific realists, we aim to understand a world that exists independently of our perceptions and communicate insights about how that world works. Based on observation and experimentation, we create and test hypotheses and construct theories that approximate this underlying reality. In doing so, we form internal models, i.e., **cognitive representations** of **real entities** and their relationships, which we externalize as data and knowledge (i.e., **representational artifacts**) to communicate with others (1,2) (see also *semiotic triangle* in (3)).

Representational artifacts take various forms. Natural language texts and structured datasets convey meaning through shared conventions regarding both the intensional meaning and reference of terms and symbols (*non-natural meaning* (4)), while images, audio-visual recordings, and virtual or physical 3D models rely on resemblance (*natural meaning* (4,5)). In both cases, but especially in the former, communication depends on the recipient's ability to reconstruct the sender's cognitive representation, which in turn requires shared background knowledge. This applies equally to human-human, human-machine, and machine-machine communication.

All data and knowledge representations are shaped by the context in which they were created and the purposes they serve. As such, they are not neutral observations but models, and thus abstractions of real-world systems that support reasoning, decision-making, and further inquiry. For the purposes of this paper, our definition and discussion of a 'model' is much broader than the sorts of models usually considered by ecologists (i.e., statistical models such as general linear models). In ecology, the challenges of constructing models in this general sense are amplified by the diversity and heterogeneity of data and knowledge sources. Data relevant to ecological questions originate from disciplines as varied as soil chemistry, climatology, animal behavior, remote sensing, and physiology, each using distinct terminologies, resolutions, and modelling conventions. Similarly, relevant knowledge can originate from different sources such as scientific publications, policy reports or local and traditional knowledge.

Ecologists frequently encounter difficulties synthesizing data and knowledge from diverse sources. Meaningful reuse of representational artifacts depends not only on shared formats but also on capturing **granularity**, **complexity**, and **context**, which often requires rich metadata and fine-grained documentation. These challenges are increasingly urgent as the volume of ecological data continues to grow. Further, ecologists frequently aim to gain causal understanding of a system: if causal relationships are known, not only can (for example) the impacts of environmental change or perturbations be predicted, but action can be taken to yield desired outcomes. However, confounding variables can bias the interpretation of causal relationships (6), a problem compounded when attempting to unite information from disparate datasets.

In response to these challenges, we advocate a **modelling perspective** in which all internal and external representations of ecological data and knowledge are understood as models, shaped by context, constrained by purpose, and interpretable through conventions. Models are not replicas but **purposeful surrogates** of a referent system[1] (7). As Stachowiak (8) outlines, models possess three key features: they map aspects of a referent system (mapping feature), they simplify by excluding irrelevant details (reduction feature), and they are designed for use in a specific context (pragmatic feature) (8,9). Understood in this way, **not every representation qualifies as a model**. A stray

---

[1] not to be confused with a reference system, which is a system that functions as a reference to one or more systems

photograph or a sensor log may represent reality, but it becomes a model only when interpreted as a representation of a particular referent system that has been modelled to serve a specific use context. Scientific models, whether expressed as data structures, diagrams, or theories and hypotheses, are therefore **intentional**, **selective**, and **context-sensitive**. It is this purposeful structuring that enables models to support reasoning, prediction, and communication, especially in complex domains such as ecology.

Internal cognitive models as well as models that are representational artifacts can vary in terms of their specificity and the nature of their referent system. For the context of semantic modelling, it is therefore helpful to distinguish between three basic types of models: Token, type and metamodels. **Token models** represent the properties and relationships of individual entities (9). Empirical data, documented using natural language or data structures, such as "*Soil sample X has a mass density of 0.57 g/cm³*," are an example for token models (3). By classifying the elements representing individual entities within token models, token models can be transformed into **type models** (e.g., "*SOIL_SAMPLE has a MASS_DENSITY of VALUE DENSITY-UNIT*"), where each element is represented by a slot (e.g., *SOIL_SAMPLE*, *VALUE*) that accepts only individual entities of a specific type (i.e., slot-constraint). By further generalizing between the classified elements, type models can be transformed into **metamodels** (e.g., "*MATERIAL_OBJECT has a QUALITY/DISPOSITION of VALUE UNIT*") (9). Scientific knowledge in the form of universally applicable theories and causal relationships, but also universally applicable class axioms (i.e., machine-interpretable class definitions), represent **universal metamodels**. Data schemata (i.e., a structured blueprint that defines how data is organized, enabling their consistent storage, retrieval, and interpretation), on the other hand, are **schema metamodels**. Token models instantiate their respective type and metamodels, while metamodels can be used to **validate** and structure the interpretation of token-level data.

When cognitive models (no matter whether they are token, type or metamodels) are turned into representational artifacts, they can take various forms. However, to allow for any kind of reuse, it is critical that they are both **human-intelligible** and **machine-interpretable**. Adhering to the **CLEAR Principle**, making data and knowledge **C**ognitively interoperable, semantically **L**inked, contextually **E**xplorable, easily **A**ccessible, and **R**eadable and interpretable for humans, is essential for their human interpretability and reusability. Similarly, adhering to the **FAIR Principles** (10), making data and knowledge **F**indable, **A**ccessible, **I**nteroperable, and **R**eusable, is essential for enabling automated reasoning, cross-domain data integration, and the modelling of causal relationships.

In this paper, we introduce a **semantic units based modelling approach** that emphasizes human-interpretability while supporting high levels of FAIRness, CLEAR-compliance, AI-readiness, and logical reasoning capability. This approach offers a granular, composable and transparent way to represent data and knowledge, making it suitable for aligning heterogeneous ecological information and supporting causal modelling. Semantic units can be serialized as **FAIR Digital Objects (FDOs)** (11,12), a format that encapsulates data, metadata, and semantics in a machine-actionable and persistently identifiable structure, further supporting their robust reuse, integration, and automated reasoning across ecological domains.

Before going into detail with our suggestion, in *Basics of Semantic Modelling* we first introduce the foundations of semantic modelling, including RDF, OWL, ontologies, and knowledge graphs, and describe the concept of semantic units. This chapter provides the technical foundation of our approach for readers interested in the concepts, methods, and principles underlying semantic modelling of FAIR and CLEAR, machine-actionable ecological data and knowledge.

In *Towards a Foundational Framework of Modelling Data and Knowledge in Ecology*, we show how semantic units can be applied to represent different types of ecological data and knowledge in a knowledge graph, demonstrating their potential as a **foundational framework for modelling ecological data and knowledge** and high-level synthesis, as well as for facilitating interoperability and causal reasoning across ecological domains. The section *Modelling Causal Relations* may be of particular interest. Here, we illustrate how individual causal hypotheses can be conceptualized as semantic units and then combined to construct causal networks. The resulting causal map serves as a semantic grid comparable to a roadmap to navigate causal relationships, their associated empirical data, and related knowledge sources. In this section, we also examine the documentation of causal junctions and the definition of particular causal perspectives to account for contextuality. These are subjects that are likely to be of interest to ecologists and practitioners.

# Basics of Semantic Modelling

In this section, we introduce key semantic technologies and concepts that form the foundation of our framework for modeling ecological data and knowledge, including causal relationships. Originally developed in the fields of computer science and artificial intelligence, these technologies and concepts allow us to express ecological knowledge and empirical data in a structured, interoperable format that supports reasoning, querying, and integration. While these methods may seem distant from fieldwork or ecological theory at first glance, they offer powerful tools for organizing, connecting, and making sense of complex ecological information.

**Note on Terminology**
The following sections include definitions and characterizations of key concepts of semantic modelling as we adopt them in this article to support the Semantic Units Framework. They are consistent and sufficient for the purpose of this work, but we acknowledge that other authors may employ different or more nuanced definitions. Our usage is designed to balance conceptual clarity for ecologists with formal precision for semantic modelling. This applies in particular to our characterization of ontologies, semantic schemata, and knowledge graphs.

## Expressing Knowledge and Data as RDF Triples

At the core of semantic modelling is the **Resource Description Framework (RDF)**, a data model for expressing information in a machine-readable form. RDF represents data and knowledge as simple **triples** with three components: a *Subject*, a *Predicate*, and an *Object*. These triples make statements about **resources**, i.e., things we want to describe.

Each resource is identified by an **Internationalized Resource Identifier (IRI)**, a globally unique string that enables consistent referencing. To avoid confusion with ecological uses of the word "resource", we refer to entities identified by IRIs as *identifier resources* (**ID resources**). These may represent a concept (e.g., '*ecosystem*', '*planting*'), an individual object (e.g., '*ecosystem 123*') or process (e.g., '*planting 456*'), or relationships (also called properties; e.g., '*has part*').

The *Subject* and *Predicate* in a triple are always ID resources. The *Object* can either be another ID resource or just a **literal**, which is a concrete value such as a string, number, or date without an IRI (e.g., '*20.5*' or '*Amazon rainforest*' as the label of a resource representing the Amazon rainforest).

Here is a basic example of an RDF triple:

[1]        S:[http://purl.obolibrary.org/obo/ENVO_01001939](http://purl.obolibrary.org/obo/ENVO_01001939),
P:[http://www.w3.org/2000/01/rdf-schema#subClassOf](http://www.w3.org/2000/01/rdf-schema#subClassOf),
O:[http://purl.obolibrary.org/obo/ENVO_01001110](http://purl.obolibrary.org/obo/ENVO_01001110)

This triple states that one concept (*'ecosystem'*) is a subclass of another concept ('*environmental system*'), using IRIs from the **Environment Ontology (ENVO)** as their identifiers and standard RDF Schema vocabulary. Since humans are not good at reading IRIs, a human-readable label can be specified for each ID resource in an RDF document by adding a triple with the respective IRI as *Subject*, '`label`' [`RDFS:label`] as *Predicate*, and the label as a literal in the *Object* position. To improve readability, in this paper we will generally use human-friendly labels accompanied by ontology acronyms and IDs in place of the IRIs:

[2]        S:'`ecosystem`' [`ENVO:01001939`],
P:'`subclass of`'[`rdfs:subClassOf`],
O:'`environmental system`'[`ENVO:01001110`]

When the ID resource in the *Object* position of one triple is shared in the *Subject* position of another triple, RDF triples become connected to form large **semantic graphs** or **knowledge graphs** and thus a network of information in which nodes (ID resources in *Subject* and *Object* positions) represent entities and concepts, and edges (ID resources in *Predicate* positions) represent relationships. These graphs allow machines to navigate, query, and reason over large quantities of structured ecological knowledge and data.

## Making Statements about Statements

The RDF triple model is well suited for expressing **binary relationships** and thus statements that relate just two entities, such as "*ecosystem* is a subclass of *environmental system*" (see example [2]). However, many ecological statements involve more than two entities, and it usually takes multiple triples to fully express them. For instance, consider an ecological intervention event in which several individuals of a plant species are planted at a specific site and date, by a particular restoration organization, and for a defined purpose. This is an **n-ary relationship**, as it involves multiple entities that cannot be fully expressed using a single RDF triple.

In practice, such intervention events often carry additional contextual information. We may want to record whether this intervention was part of a specific field study, whether it was planned based on expert judgment, or what outcome it aimed to achieve. In other words, **we want to make statements about n-ary statements**.

One solution to this problem is the use of **Named Graphs**. A Named Graph is a set of RDF triples that are grouped together and assigned their own ID resource. This ID resource allows us to refer to the entire group of triples as a single unit.

To implement this, RDF triples are extended into **quads** by adding a fourth element: the graph name. A quad takes the form:

[3]        Subject, Predicate, Object, GraphName

Here is an example for a Named Graph comprising four triples:

[4]        Named Graph:'**`terrestrialEcosystem_123`**':

```
[4.1]       S:'ecosystem_123',
            P:'type'[rdf:type],
            O:'terrestrialEcosystem'[ENVO:01001790],
            GN:'terrestrialEcosystem_123'

[4.2]       S:'ecosystem_123',
            P:'hasPart'[BFO:0000051],
            O:'forestedArea_123'[ENVO:00000111],
            GN:'terrestrialEcosystem_123'

[4.3]       S:'forestedArea_123'[ENVO:00000111],
            P:'participatesIn'[RO:0000056],
            O:'plantingProcess_456'[AGRO:00000231],
            GN:'terrestrialEcosystem_123'

[4.4]       S:'plantingProcess_456'[AGRO:00000231],
            P:'hasPart'[BFO:0000051],
            O:'irrigationProcess'[AGRO:00000006],
            GN:'terrestrialEcosystem_123'
```

Translated into plain language, this Named Graph describes a concrete ecosystem ('terrestrialEcosystem_123'), telling that it is a terrestrial ecosystem [4.1] that is partly forested [4.2], and that in that specific forested part ('forestedArea_123') something is being planted [4.3], which involves irrigation [4.4].

All quads that share the same GraphName (=ID resource; 'terrestrialEcosystem_123' in our example) are understood to belong to the same Named Graph. We can then use this GraphName ID resource in other triples to annotate or describe the entire group, for example, linking it to a data source, method, or confidence level. For example, we could tell that a scientific study happens in the above ecosystem:

```
[5]         S:'terrestrialEcosystem_123',
            P:'participatesIn'[RO:0000056],
            O:'studyDesignExecution789'[OBI:0000471]
```

Similarly, it could be expressed that a particular observation supports or contradicts a causal hypothesis, with the observation and the hypothesis being modelled each in their own Named Graphs.

This approach thus allows to represent n-ary relationships and to attach provenance, context, and meta-level information to sets of RDF triples, which is an essential capability for modelling ecological interventions, field observations, or experimental designs.

# Adding Semantics: Description Logics and the Web Ontology Language (OWL)

While RDF provides a syntactic framework for describing data and knowledge, the **Web Ontology Language (OWL)** provides a **language** that adds formal semantics grounded in Description Logics, a family of formal logics used to represent structured knowledge. OWL has been developed for the purpose of constructing ontologies, which will be described below, and can be represented (i.e., serialized) using the RDF triple syntax of *Subject*, *Predicate*, and *Object*. Using OWL enables the specification of precise, computable models of ecological knowledge and empirical data. Since it is based on Description Logics, OWL describes the world using three primary constructs: classes, instances, and properties, each of which represents a specific category of ID resources. **Classes** (or *concepts*) represent general categories of things, such as `'ecosystem' [ENVO:01001939]` or `'biodegradation' [ENVO:06105014]`, and are used to define sets of entities with shared characteristics. **Instances** (or *named individuals*), refer to particular entities, such as a given individual ecosystem (*terrestrialEcosystem_123*), an individual wolf (*wolf_A17*), or a field observation event (*fieldObEv_789*). In OWL, a given instance instantiates a target class and at the same time the technical OWL class `'named individual' [OWL:namedIndividual]` to clearly indicate that this ID resource is an instance of the target class. A class can have many instances, and an instance can belong to multiple classes (e.g., a particular wolf `'wolf_123'` can be an instance of `'Canis lupaster' [NCBITAXON:1707807]`, but also of `'carnivor' [ECOCORE:00000088]`).

Contrary to RDF, OWL clearly distinguishes three categories of properties (i.e., binary *relations*): **Object properties** connect instances to other instances via binary relationships, such as `'has part' [BFO:0000051]` or `'quality of' [RO:0000080]`, **data properties** connect an instance with a literal, such as `'has measurement value' [IAO:0000004]`, and **annotation properties** can connect classes or instances to other classes, instances, or literals. They are usually used for lexical statements and thus statements about linguistic entities, such as `'definition' [IAO_0000115]` for giving a human-readable definition for a concept or `'label' [RDFS:label]` for providing a concept's label.

Based on these three primary constructs, OWL distinguishes between two levels of information that can be modelled as interconnected RDF triples:

- The **TBox** (Terminological Box) defines the conceptual vocabulary: it includes class hierarchies (e.g., `'swampArea' [ENVO:01001208]` is subclass of `'wetlandArea' [ENVO:00000043]`) and formal constraints (e.g., `'wetlandArea' [ENVO:00000043]` is equivalent to a `'vegetatedArea' [ENVO:01001305]` that `'overlaps' [RO:0002131]` with some `'wetlandEcosystem' [ENVO:01001209]`). Universal metamodels are thus represented within the TBox.
- The **ABox** (Assertional Box) contains assertions about individuals, such as `'terrestrialEcosystem_123'` is `'locatedIn' [RO:0001025]` `'wetland_A'` or `'bumblebee_X'` `'pollinates' [RO:0002455]` `'AsterBreviscapus_Z'`. Token models are thus represented within the ABox.

A major strength of Description Logics is that they enable **automated reasoning**. OWL-compatible reasoners can infer implicit facts, validate data, and detect contradictions. Description Logics support a variety of reasoning tasks:

- **Class subsumption**: For example, determining that all terrestrial ecosystems are ecosystems based on the TBox definitions of the two classes;
- **Instance checking**: For example, verifying whether a given site qualifies as a terrestrial ecosystem based on its ABox assertions and the ecosystem type's TBox definition;

- **Consistency checking**: Identifying logical conflicts, such as whether ABox assertions exist that contradict TBox definitions.

These reasoning capabilities have the potential to help ecologists to infer new facts, validate models, uncover hidden relationships and explore cascading effects in complex ecological systems. By combining formal class definitions with instance-level data, Description Logics have the potential to enable scalable, interoperable representations of ecological knowledge, integrating alternative terminologies and classifications, interaction networks, land-use types, and field observations into coherent, queryable knowledge systems. Understanding how classes, instances, and object properties interact in OWL lays the groundwork for building ontologies.

## OWL Ontologies as Terminologies Based on Formal Semantics

OWL allows the definition of controlled vocabularies based on formal semantics, known as ontologies. Ontologies are a core building block of semantic modelling. An **ontology** is a formal specification of a shared conceptualization, i.e., a structured, logically grounded representation of knowledge within a particular domain, such as ecology. In practice, an ontology consists of a vocabulary of key domain concepts, represented as classes, relationships between them described by object and data properties, and axioms that constrain how those classes and properties can be combined. These elements are typically organized in taxonomies (not to be confused with Linnean taxonomy) via subclass or subproperty hierarchies, which help clarify domain structure and support automated reasoning.

For example, in restoration ecology, we might define an ontology in which `'habitat'`, `'species'`, and `'restorationActivity'` are classes. We might then assert that every `'restorationActivity'` `'targets'` some `'habitat'` and `'affects'` some `'species'`, implying that this relation **necessarily holds true for every instance** of `'restorationActivity'`. Ontologies are therefore collections of such universal metamodels. By encoding these relationships formally, we enable consistent interpretation and integration of data and knowledge across studies, projects, and geographic regions.

It is important to distinguish between the **semantics** of an ontology and its **syntax**. Semantics refers to the meaning of the ontology's constructs: how classes, properties, and instances are logically interpreted, typically using Description Logics. Formal semantics allows automated reasoners to derive logical consequences of a given set of axioms. Syntax, by contrast, refers to how this information is expressed or serialized in a machine-readable format. Common syntaxes include RDF/XML, Turtle, and JSON-LD. Each of these provides different ways to encode the same semantic content, and choice of syntax often depends on tooling or interoperability needs rather than any impact on reasoning or expressivity.

## Semantic Schemata are the Building Blocks of FAIR Knowledge Graphs

Ontologies define the foundational concepts of a domain in the form of universal statements, typically taking the form of OWL class axioms (the TBox). These axioms represent universal metamodels, specifying universally applicable statements about the domain. In contrast, **knowledge graphs** focus on individual instances and their relationships, capturing instance-level information (the ABox) including empirical data, case-specific assertions, and observations. These reflect token models.

To ensure that ABox-level assertions remain consistent, interpretable, and interoperable, they must be structured according to shared modelling conventions. This is the role of semantic schemata. A **semantic schema** provides a design pattern for representing a specific type of information in a knowledge graph using one or more triples. It specifies which ontology classes and properties should be used, how entities relate to one another, and what constraints or expectations must be respected. As such, a semantic schema functions as a **schema metamodel**: a specification of how token models should be constructed based on the domain's universal metamodels.

Semantic schemata are typically formalized using the **Shapes Constraint Language (SHACL)**. SHACL enables the definition of *shapes*, i.e., structured templates that describe the expected form of RDF data. For example, a SHACL shape might specify that every '`restorationActivity`' must '`target`' some '`habitat`', or that an '`observationEvent`' must include a '`date`' and a '`location`'. These constraints ensure that data from different sources can be validated against a common semantic pattern before being integrated into a knowledge graph.

Importantly, semantic schemata draw directly on existing ontologies. By referencing OWL classes and properties, they help unify empirical data and other sources of knowledge with logically defined domain knowledge. This combination makes it possible to build graphs that are not only **machine-readable**, but also **machine-actionable**, allowing software to validate structure, enforce constraints, and even infer new facts.

In this way, semantic schemata serve as the bridge between abstract knowledge models and concrete data, forming the **building blocks of FAIR knowledge graphs**. While ontologies provide the logic and vocabulary for describing a domain, semantic schemata ensure that data instantiated within that domain adhere to shared semantic expectations. This is especially critical in ecology, where data are often collected under diverse protocols and vocabularies. By using schemata to guide data modelling, we promote **semantic interoperability**: the ability of systems and users to understand and process data meaningfully, even when it originates from heterogeneous sources.

While ontologies and semantic schemata provide the backbone of structured, logically coherent knowledge graphs, they primarily operate at the level of classes, shapes, and broad modeling patterns. However, ecological data and knowledge often require finer-grained structuring into specific causal assertions, individual field observations, or particular management interventions that frequently span different logical formalisms, granularities, or epistemic statuses. For such cases, schema-level models alone are not sufficient to capture the nuance and reusability required. To address these challenges, we introduce a more modular strategy for representing and managing ecological knowledge: the **Semantic Units Framework**, which builds on the foundation of semantic schemata by decomposing data and knowledge into independently meaningful and logic-aware units of information.

## The Semantic Units Framework

While OWL-based knowledge graphs offer a powerful foundation for representing structured ecological knowledge, they also present notable limitations. OWL's design prioritizes logical rigor but restricts the kinds of statements that can be expressed and introduces modelling complexity that often exceeds what is practical for domain experts—we address some of these limitations in the following sections. As a result, many users, especially ecologists and applied researchers, find semantic technologies difficult to adopt in day-to-day data management.

To address these challenges, we will now introduce the **Semantic Units Framework** in two steps. First, we will introduce the foundational principle of **semantic modularization**. Second, we will introduce the concept of Semantic Units with its two major types, statement units and compound units, and will show how semantic units can be serialized into FAIR Digital Objects.

## The Semantic Modularization Principle

Restoration ecology requires a wide spectrum of ecological data and knowledge from diverse sources, from detailed field observations to empirical generalizations, community-based knowledge and theoretical background knowledge. These forms of knowledge vary in precision, context, and abstraction levels, and are often highly interdependent, with a nested granularity. OWL, as a language for representing knowledge graphs, alone cannot capture this heterogeneity and granularity without becoming either overly complex or semantically rigid. To reconcile these tensions, the semantic units approach applies the **semantic modularization principle** (13): decompose ecological knowledge and data into self-contained units that are logic-aware (i.e., indicate which logical framework has been applied for the semantic modelling of the unit itself; e.g. Description Logics or First-Order logic) and independently addressable (i.e., each unit possesses its own ID resources). In this way, the units preserve meaning, support reuse, maintain cognitive accessibility, and can be annotated, reasoned over, queried, or ignored, depending on the task at hand.

Semantic modularization has five key benefits:

- **Logical heterogeneity**: different semantic units can use different logic formalisms (e.g., semantic content modelled in OWL uses Description Logics, but other approaches, such as Common Logic (ISO/IEC 24707:2018), use other logical frameworks, such as First-Order Logic) that are best suited for modelling their content;
- **Semantic coherence**: each unit maintains internal consistency, with its content being meaningful to a human user;
- **Cognitive interoperability** (Box 1): units are comprehensible and usable by diverse users;
- **Contextual navigability**: units can be linked to other relevant units and support contextual explorability of a knowledge graph (14);
- **Nesting**: units can be nested within units, referenced, or combined across the knowledge graph.

**Box 1 | Cognitive Interoperability** (14)

Cognitive interoperability refers to a system's ability to support intuitive and efficient human interaction with data and metadata. It focuses on aligning the complexity of data structures (human-information interaction) and interfaces (human-computer interaction) within human cognitive capacities. Systems that support cognitive interoperability offer tools for exploration, trust building, and integration into workflows, making them accessible to database architects, data scientists, and domain experts alike.

## Semantic Units

Semantic units (15) comply with the semantic modularization principle and are **modular subgraphs** consisting of one or more triples that represent a discrete unit of information, i.e., the unit's **semantic content** (13). A semantic unit's content is always **semantically meaningful to a human user**, forming a semantically complete statement or a collection of such statements (for a counter example, see the single triple highlighted in red in Fig. 2B,C). Each semantic unit is represented in the graph with a globally unique and persistent ID resource. This ID resource functions as an identifier for the semantic content the unit carries. This is achieved by organizing the semantic content of the unit in a Named Graph, forming the **content-graph** of the unit. It contains all triples modelling the semantic content and provides a single IRI, i.e., its ID resource, that simultaneously is the ID resource of the semantic unit itself (indicated by the blue arrow in Fig. 1) and thus can be used in the content's place. This ID resource instantiates a corresponding semantic unit class, which in turn classifies the semantic content to be a specific type of information. The ID resource allows the unit to be referenced, annotated, and reused independently, for instance, to specify the content's source, which could be a paper, report, or dataset. This information, and other types of metadata, including provenance and licenses, is captured in the **meta-graph** of the unit (Fig. 1).

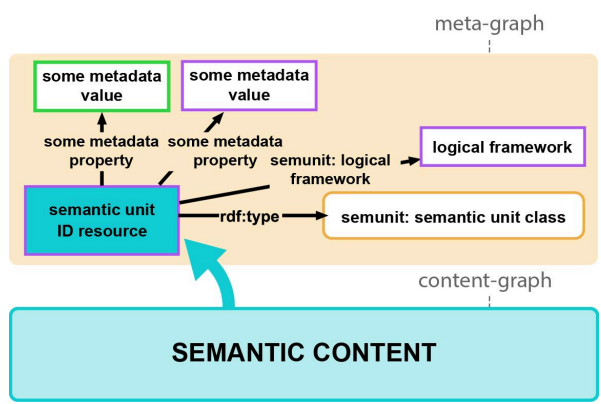

**Figure 1: Basic structure of a semantic unit, implemented in an OWL/RDF framework.** A semantic unit is structured into two parts: At the bottom, the content-graph contains the unit's semantic content (here only shown as a placeholder for the actual graph; cf. with Fig. 2C); at the top, the meta-graph contains the unit's metadata. Metadata may include provenance, reference to the SHACL shape that has been used for modelling the semantic content, a license specification, and the specification of the logical framework being used for modelling the content. The semantic unit ID resource (blue box with purple border) is an identifier for the unit's semantic content and thus the unit's entire content-graph. The unit's ID resource instantiates a semantic unit class, indicating that the unit's semantic content is of a specific content type. *Class ID resources shown with yellow border, instance ID resources with purple border, and literals with green border.*

Semantic units are organized into two primary categories, each tailored to different representational needs: statement units and compound units.

### Statement Units

Statement units (13,15) are the **atomic building blocks** of semantic representation. They are the smallest units of semantically meaningful content. Each statement unit captures a single proposition, such as "*Soil sample X has mass density of 0.57 g/cm³*" (Fig. 2A), and organizes it as a self-contained RDF/OWL subgraph (Fig. 2B), using ID resources in the subject position and either ID resources (i.e., resource-objects) or literals (literal-objects) in the object positions. Organized in a Named Graph, classified as an instance of a corresponding statement class, and accompanied by respective metadata, it forms a statement unit (Figure 2C). Each statement unit adheres to a SHACL shape that defines its structure and ensures it can be interpreted and queried consistently, supporting the overall interoperability and machine-actionability of its semantic content.

Linked to the SHACL shape, templates provide ways to dynamically display the unit's semantic content as human-readable and -interpretable text or graph (see *dynamic label* and *dynamic graph* in

Fig. 2 D,E), supporting their **cognitive interoperability**. Consequently, users of the knowledge graph do not have to understand the underlying data structure of a semantic unit, which often includes triples that do not make sense to a domain expert. This is important, as to make ecological data and knowledge machine-actionable, models that support automated reasoning and semantic interoperability are often highly formalized, multi-layered, verbose, and technically complex, utilizing terms from many ontologies, and are dense with formal syntax (e.g., IRIs, restrictions, reifications). While machines can process such structures efficiently, humans often struggle with correctly interpreting them without training or the help of specialized tools (cf. triple highlighted in red in Fig. 2C). The dynamic labels and graphs strengthen the cognitive interoperability of the semantic units.

## A) Statement

*Soil sample X has a mass density of 0.57 g/cm³*

## B) OWL/RDF Graph

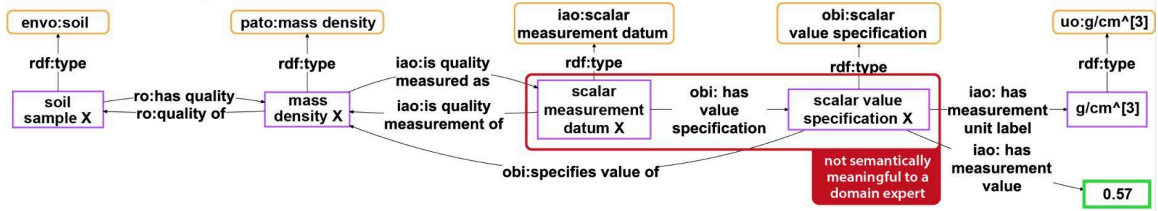

## C) Statement Unit

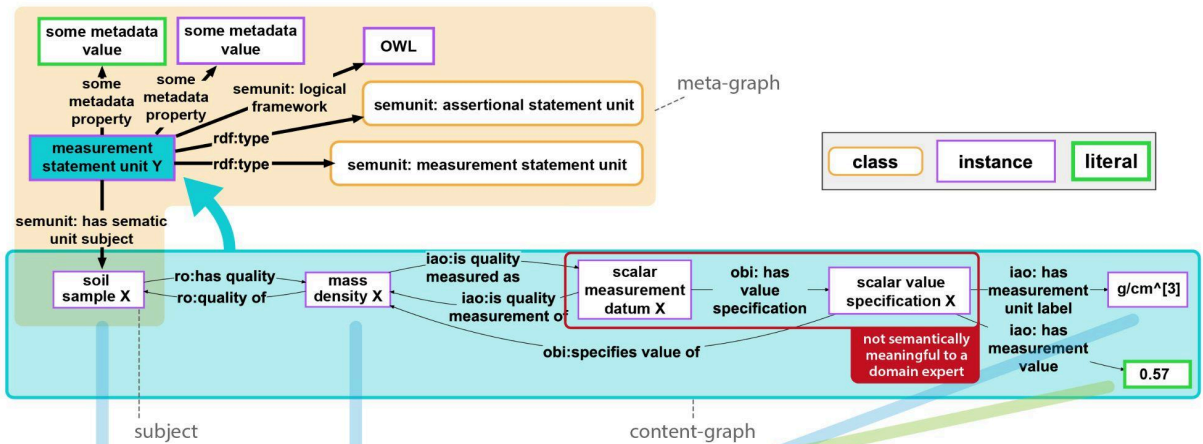

## D) Dynamic Label (Textual Display)

**'soil sample X has a density of 0.57 g/cm³'**

## E) Dynamic Graph (Graphical Display)

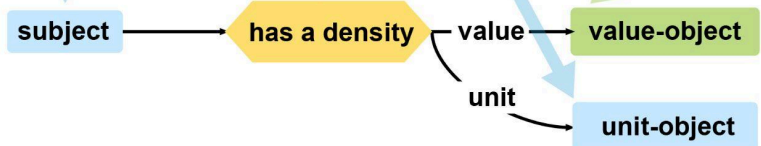

**Figure 2**: **From natural language statement to OWL graph to semantic unit**: **A)** A human-readable assertional statement about the mass density of a specific soil sample. **B)** The representation of the statement from A) as a semantic graph, adhering to RDF syntax and following an established semantic schema for modelling measurement data from the Ontology of Biomedical Investigations (OBI). **C)** The same graph, organized as a statement unit. The information from A) and B) is contained in the content-graph, denoted within the blue-bordered box, articulating the statement from A), modelled as in B). The unit's meta-graph is shown in the peach-colored box, denoting the unit's ID resource (blue box with purple border) as an instance of the '`measurement statement unit`' class and the '`assertional statement unit`' class. The statement unit ID resource has the same identifier as the unit's content-graph, and thus can be used in place of the unit's semantic content (indicated by the blue arrow). The peach-colored box also contains various metadata triples, here only indicated by 'some metadata property' and 'some metadata value' as their placeholders. They may include provenance, reference to the SHACL shape that has been used for modelling the semantic content, and a license specification. Part of the metadata is also the specification of the logical framework that was applied for modelling the content. **D,E)** The dynamic label and dynamic graph associated with the statement unit for a human-interpretable display of the unit's content.

## Compound Units

Compound units (13,15) enable the modelling of hierarchical levels of organization, allowing the granular nesting of semantic content, by organizing statement units and finer-grained compound units into semantically coherent and meaningful collections (Figure 3). As a semantic unit, every compound unit possesses its own ID resource that instantiates a corresponding compound unit class. Contrary to statement units, a compound unit does not possess its own semantic content, but receives this via directly or indirectly (through an association-chain of two or more compound units levels) associated statement units. A compound unit may for instance group all statement units about a particular field study, species, or ecosystem restoration case.

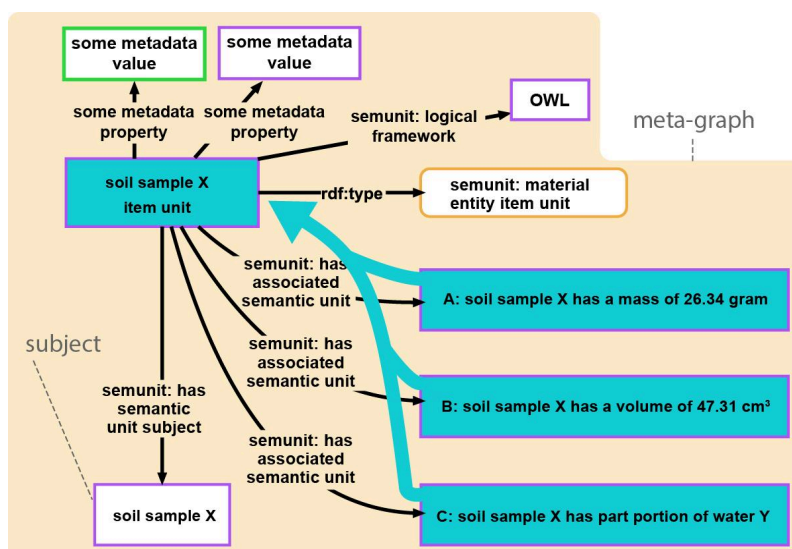

**Figure 3: Compound unit example.** The compound unit ID resource is denoted as '`soil sample X item unit`' and instantiates the '`material entity item unit`' class. It encompasses all statement units that share '`soil sample X`' as their subject, and it therefore provides a description of '`soil sample X`' (the associated statement units A, B, and C describe different characteristics of the same soil sample). Compound units do not possess their own content-graph and thus do not directly carry semantic content, but obtain it by referencing the ID resources of their associated semantic units. One can obtain their respective content-graphs and merge them to form the semantic content of the compound unit (here indicated by the blue arrows). Each compound unit is represented in the knowledge graph by its compound unit ID resource and all triples in its meta-graph (depicted in the peach-colored box), referencing the ID resources of their associated semantic units (here: A, B, and C are shown with their dynamic label instead of their ID resource IRIs, for human-readability of the figure).

## Serializing Semantic Units to FAIR Digital Objects

The modularity and semantic clarity of semantic units make them ideal candidates for structured, persistent, and interoperable digital knowledge artifacts. Each semantic unit, whether representing a simple assertion or a compound structure, is independently identifiable, logic-aware, and enriched with context. However, to fully leverage these features, especially for sharing, referencing, and computational reuse across systems, we require a mechanism that preserves not just the data but its semantic intent, structure, and context. This is precisely where the concept of **FAIR Digital Objects**

**(FDOs)** (11,12) becomes relevant. It provides a formal infrastructure for turning semantic units into self-contained, findable, accessible, and machine-actionable components of ecological knowledge.

While we focus in this paper on applying semantic modelling to ecological data and knowledge for representing them in knowledge graphs, the semantic units approach is in principle technology-independent and can be implemented in different database frameworks, including relational databases. Here, statement units would be organized in tables specific for each statement unit class, where the ID resource of a unit would be specified in a cell of a row that contains the unit's semantic content (13).

Depending on the technical implementation, different approaches for serializing semantic units into **FAIR Digital Objects (FDOs)** exist. Statement units documented in a **knowledge graph**, for instance, can be serialized as **Nanopublication FDOs** (14,15). A nanopublication is a self-referential collection of defined Named Graphs, with an Assertion Named Graph containing the semantic content to be published and a Provenance and a Publication Named Graph containing metadata about the content and about the nanopublication itself, respectively. A Header Named Graph lists the ID resources of these three Named Graphs and thereby connects them to a single nanopublication. Nanopublications have been used before for documenting ecological hypotheses (see, e.g., 'adaption in response to enemy release in the non-native range increases population-level performance of non-native species' in (16)).

As such, a nanopublication is an autonomous and self-contained publication that possesses its own DOI and qualifies as a FDO. It can be used to publish individual assertions and thus allows a much finer-grained publication than the typical scholarly publications. Nanopublication FDOs are published in a format that is both formal and machine-interpretable.

Compound units can be serialized as **nested Nanopublication FDOs**, which differ from statement Nanopublication FDOs in that their Association Named Graph is empty, and their Header Named Graph lists additionally the DOIs of the statement Nanopublication FDOs that comprise the compound unit.

Statement units documented in a **relational database**, on the other hand, can be serialized as **Research Object Crates (RO-Crates)**. An RO-Crate is a Research Object (RO) that comprises a collection of data structures (Crate), with an RO-Crate-metadata.json file describing the collection. RO-Crates can comprise a variety of types of research data and are not limited to RDF/OWL-based data. They can include PDFs, entire datasets, software, and references to other research, even to Nanopublications. For an overview of the key concepts in the semantic units approach, see Table 1.

**Table 1: Overview of key concepts in the semantic units approach**

| Concept | Description |
|---|---|
| **Semantic Unit** | A logic-aware, context-rich, modular unit of (ecological) data or knowledge. |
| **Named Graph** | An RDF construct used to group triples under a shared ID resource (i.e., IRI). |
| **FAIR Digital Object** | A digital artifact with persistent ID, metadata, and defined operations. |
| **Semantic Unit as FAIR Digital Object** | A semantic unit serialized as a FAIR Digital Object for reuse and sharing. |

# Towards a Foundational Framework of Modelling Data and Knowledge in Ecology

After having introduced the basic idea of semantic units, we will now describe how semantic units can be utilized for representing different types of ecological data and knowledge, and how to combine them in a knowledge graph. The overall aim of this chapter is to suggest a **foundational framework for modelling data and knowledge in ecology**. We believe that this framework will create interoperability, will facilitate causal reasoning across ecological domains, and therefore has the potential to strongly enhance synthesis for ecological theory and practice. We will start by focusing on how to model different categories of statements before we address the modelling of causal relations.

## Ecological Data and Knowledge Involves Different Statement Categories

Ecologists communicate data and knowledge using different categories of statements. Some statements describe what has been directly observed or measured in a particular place and time, while others express broader, general patterns or rules, still others define how we organize and describe ecological entities themselves.

To model ecological data and knowledge in a way that is both machine-actionable and human-interpretable, and thus FAIR and CLEAR, it is important to distinguish between these categories. This distinction affects how statements are represented, connected to evidence within the knowledge graph, combined into larger compound units, and reused across studies. In this section, we introduce the main statement categories used in the Semantic Units Framework, discuss some modelling limitations of OWL and how the framework solves the limitations, and explain their roles in ecological knowledge representation.

### Assertional and Universal Statement Units

At the heart of ecological data are assertional statements, communicating specific claims about particular observations, measurements, or events that occurred at a given place and time. In contrast, at the heart of ecological knowledge are universal statements, communicating general claims or rules that apply across many cases, capturing patterns, processes, or relationships thought to hold true within a domain. In the Semantic Units Framework, assertional and universal statement units form the two core building blocks that connect concrete empirical evidence to broader theoretical understanding, including specific hypotheses and theories.

As introduced above, the language OWL supports three basic types of ID resources: classes, instances, and properties. Together, they allow for modelling two fundamental types of statements:

- **Assertional statements** describe concrete facts about individual entities. For example, a statement such as "*Soil sample X has a mass density of 0.57 g/cm³*" (cf. Figure 2A) applies only to that specific entity. It may or may not be true in other contexts and for other entities. These statements represent **token models** and are encoded in OWL using **instance ID** **resources**. The resulting triples belong to the **ABox**. In the semantic units approach, such assertions are modelled as **assertional statement units** (see Fig. 2C). Examples include qualitative descriptions of individual material entities and their parts and properties, qualitative descriptions of individual processes and their order of events, inputs, outputs, and participants, and quantitative measurement data and other observational data.

- **Universal statements** express what is **necessarily** true for all instances of a given class. For example, "*Every wetland area contains some wetland ecosystems*" expresses a generalization about the class 'wetlandArea' [ENVO:00000043] (contains would be expressed using 'overlaps' [RO:0002131]). These statements represent **universal metamodels**, modelled in OWL through class axioms using **class ID resources**, and belong to the **TBox**. Examples include established domain knowledge, such as universally applicable class axioms, but also scientific laws and hypotheses, including causal relationships, which can only be modelled as class axioms in OWL.

Both types of statements are **truth-apt**, i.e., they make direct truth claims that can be evaluated as true or false. However, OWL's TBox-ABox distinction presents a challenge: universal statements are encoded as class axioms but do not have their own ID resources (i.e., class axioms are documented within the corresponding class specification—referencing them using that class' ID resource would introduce ambiguity, since that ID resource refers to the entire class concept and not only its class axiom(s)). As a result, individual class axiom expressions remain inaccessible to referencing and querying within a knowledge graph. They lie **outside the proposition space** of the graph—they do not belong to the set of things that the graph can talk about. In other words, **we cannot document ecological knowledge as universal statements in OWL-based knowledge graphs. Knowledge graphs would be restricted to document only assertional statements**.

To address this, we propose extending the ID resources types used in semantic modelling by introducing **every-instance** and **some instance ID resources** (13):

- A **some-instance ID resource** represents one or more unspecified instances of a class *C*. When used in a triple, this resource reads: "*There exists at least one instance of class C such that…*". To specify a some-instance ID resource, the resource has to instantiate both the target class *C* and the 'some-instance resource' class.
- An **every-instance ID resource** denotes universal quantification over a class C, indicating that a given property or relation holds for every instance of the class, individually. It reads: "*For every instance of class C, it holds that...*". To specify an every-instance ID resource, the resource has to instantiate both the target class *C* and the 'every-instance resource' class.

For example, the universal statement "*Every wetland area overlaps with some wetland ecosystem*," the example we already discussed further above, that in OWL can only be modelled as class axiom belonging to the TBox and thus outside the proposition space of a knowledge graph, can now be modelled in triple form as part of the ABox:

[6]     S: every-instance of 'wetlandArea' [ENVO:00000043],
        P: 'overlaps' [RO:0002131],
        O: some-instance of 'wetland ecosystem' [ENVO:01001209]

These extended types allow universal statements to be expressed as **universal statement units, thus becoming part of the things that a knowledge graph can talk about** (Fig. 4). Crucially, each such unit is addressable and can be referenced, annotated, or linked to empirical evidence. These units can be translated into OWL-compatible axioms, enabling automated reasoning over them (for details and discussion, see (17)).

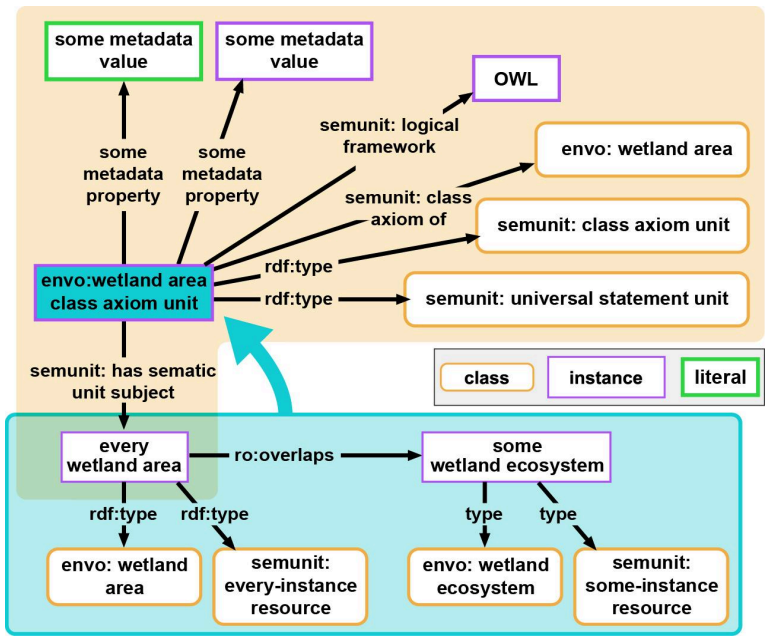

**Figure 4: Universal statement unit example.** The documentation of the universal statement from [6] as a universal statement unit. Since the semantic content of this unit specifies the class axiom for `'wetland area'[ENVO:00000043]` (indicated through the property `'class axiom of'`), it not only instantiates the class `'universal statement unit'` but also `'class axiom unit'`. Notice the use of `'every-instance resource'` and `'some-instance resource'` for modelling the unit's semantic content, i.e., the graph in the unit's content-graph (blue-bordered box).

By linking a universal statement unit to the class it defines (e.g., via a `'class axiom of'` property), we introduce axioms as explicit, addressable ABox entities, thereby proposing a clean and ontologically coherent way to bridge OWL's TBox axioms with ABox representations. Class axioms themselves become addressable and referencable through their ID resources, to which provenance, certainty, authorship, trust, or alternatives can be attached. The importance of this will become more evident when we discuss the modelling of causal relationships and the specification of causal maps, which would not be possible without universal statement units being documented within the knowledge graph.

We basically introduce a **semantic lifting**, where the axiom becomes a named ABox individual that *represents* a TBox-level semantic pattern, thereby providing a clear bridge from OWL's logical abstractions to FAIR and CLEAR, referencable, modular representations within knowledge graphs. This addresses a longstanding limitation in OWL: the inability to refer to or manage individual axioms outside of ontologies.

Now, let us consider the following example as the semantic content of a corresponding assertional statement unit:

[7]      S:`'wetlandArea_123'[ENVO:00000043],`
          P:`'overlaps'[RO:0002131],`
          O:`'wetland ecosystem_456'[ENVO:01001209]`

When comparing [7] with [6] and considering that [6] is a universal metamodel and [7] a corresponding token model, we can argue that [7] instantiates [6]. In other words, the semantic content of an assertional statement unit can instantiate the semantic content of its corresponding universal statement unit. Such relations of instantiation between assertional and universal statements can be documented within a knowledge graph by linking the ID resources of the assertional statement unit via a property `'satisfies'` with that of the corresponding universal statement unit. This relation is logical and not just structural or compositional and would allow making individual truth claims (assertional statement units) traceable back to generalizable knowledge (universal statement units), enabling meta-queries such as: "*Find all assertional statement units that instantiate this universal statement unit*" or "*Which universal statement units are supported (or contradicted) by evidence (aka: assertional statement units)?*". If measurements and evidence are linked to these assertional statement units, ecological hypotheses and theories could be linked to underlying empirical evidence via these assertional statement units (see discussion further below). It would also support model validation (do all observed assertional statements comply with their corresponding universal statements?) and axiom testing (do real-world token models confirm the general rule?), and would provide a powerful bridge between facts and generalizations, token and type or metamodels, and data and knowledge.

## Contingent and Prototypical Statement Units

Description Logics, and with them also OWL, are designed to express only statements that are universally true or false. This makes them unsuitable for modelling default reasoning, statistical generalizations, or exceptions and context-sensitive patterns, all of which are common in ecology and other fields.

Many ecological statements are **contingent** (i.e., context-dependent): they may hold in some contexts and conditions, but not universally (18). Contingent statements are often backed by empirical evidence from one or a few assertional statements but cannot be universally generalized. For example, "*Prescribed fire can promote native species richness in prairie ecosystems*" may be true under specific conditions (e.g., moderate intensity, early spring) but not in others. This kind of proposition cannot be expressed directly in OWL due to its lack of support for modal or probabilistic logic.

Using the semantic units approach combined with the newly introduced some-instance ID resource, we can express contingencies and thus that some statements are neither universally true, nor only true in one specific case. Maybe we want to add to the universal statement from [6] that "*Wetland areas can overlap with temperate ecosystems*", which can be modelled as a contingent statement unit that has the following semantic content (13):

[8]     S: some-instance of `'wetlandArea'` [ENVO:00000043],
        P:`'overlaps'` [RO:0002131],
        O: some-instance of `'temperate ecosystem'` [ENVO:01001931]

This **contingent statement unit** indicates that *at least one* wetland area overlaps with a temperate ecosystem, without claiming universal truth. Since it uses some-instance ID resources in both subject and object positions, it escapes the truth-functional constraint of OWL and can exist within a knowledge graph's propositional space—with contingent statement units we can add contingent statements to a knowledge graph, which would not be possible in a conventional OWL-based knowledge graph. Logical reasoning over such units requires, however, an external translation framework (cf. (13)).

A key modelling rule follows that algorithms can apply to automatically add corresponding contingent statements to a knowledge graph based on its assertional statements :

[9]   **Rule**:   *If an assertional statement is true, the corresponding contingent statement is also true (i.e., an observed fact confirms possibility).*

Consequently, observing a single wetland area to overlap with a temperate ecosystem would be sufficient evidence for proofing the contingent statement in [8].

The next category, i.e., **prototypical statements**, expresses what is **typically true**, though not universally. They describe empirical regularities, such as "*Most pioneer species exhibit fast growth rates.*" and are a subset of contingent statements, but more specific, since the statement is true for most instances of a class and not just for some. However, prototypical statements are not strictly truth-apt and are **tolerant of exceptions**. They reflect general patterns, empirical trends, or default expectations (i.e., they tolerate exceptions).

To model them, we introduce another ID resource type: A **most-instances resource** represents a typical but not universal subset of instances of class *C*, reading: "*For most instance of class C holds that...*". To specify a most-instances ID resource, the resource has to instantiate both the target class *C* and the '`most-instances resource`' class.

With this addition, we can model the prototypical statement from above as a prototypical statement unit that has the following semantic content (13):

[10]     S: most-instances of '`pioneerSpecies`',
         P:'`has disposition`'[RO:0000091],
         O: some-instance of '`fast growth rate`'

**Prototypical statement units** combine a most-instances ID resource in the subject position with some-instance ID resources in the object positions. Like contingent statement units, they lie outside of OWL's direct expressive scope and require translation to enable logical reasoning (13).

Two modelling rules further define their logical relationships:

[11]   **Rule**:   *If a prototypical statement is true, the corresponding contingent statement is necessarily true, since what is true in most cases also describes a possibility.*

[12]   **Rule**:   *If a universal statement is true, the corresponding prototypical and contingent statements are also true (this follows from deduction).*

Together, these enable a cascading logic of inclusivity across statement categories, from the universally necessary, to the generally true, to the possibly true. This kind of graded expressivity is essential in ecological modelling, where absolute rules are rare, and nuanced regularities dominate.

Importantly, the Semantic Units Framework provides a principled way to incorporate these statement types into a shared proposition space (i.e., the things a knowledge graph can talk about), rather than excluding them. It thus supports a more realistic and flexible modelling of ecological data and knowledge.

## Granularity in Semantic Modelling: Representations and Referents

The semantic units approach supports **layering different semantic roles** onto a single instance ID resource through **dual typing** (i.e., the ID resources instantiates more than one class), enabling both a **structural** and a **domain-specific** interpretation of the same ID resource. This dual interpretability becomes especially relevant when modelling complex ecological entities, such as individual organisms or ecosystems.

In OWL, an instance ID resource can instantiate multiple classes, allowing for the representation of overlapping semantic roles. For example, a compound unit comprising a collection of assertional statement units that describe an organism's anatomical parts, their properties, and the underlying hierarchy of part-whole relationships, can be typed both as an '`organism description compound unit`' and as an '`organism`'[OBI:0100026]. This **dual typing** is logically consistent as long as the involved classes are not declared disjoint, and it reflects two compatible interpretations of the same ID resource: one as a **structured semantic container**, the other as a **referential proxy**.

From a modelling perspective, the compound unit serves as a **semantic envelope**, i.e., a modular, logic-aware artifact that encapsulates fine-grained, contextualized assertions. Yet because it comprehensively describes a particular organism, it can also **stand in for** the organism and function as its semantic proxy. In this way, the compound unit acquires **referential identity**: its ID resource not only **represents** the structure of the model but also **refers to** a real-world entity. Consequently, the compound unit can be queried and reasoned over both as a model and an instance of '`organism`'[OBI:0100026].

This dual role mirrors a foundational distinction in semantics between intensional and extensional meaning (3). The **intensional meaning** (based on Frege's *sense* (19,20)) of a semantic unit corresponds to its internal structure and thus the set of statements it comprises and the way it encodes data and knowledge about its subject. The **extensional meaning**, by contrast, pertains to the real-world entity it denotes, such as a specific organism or ecosystem. Through dual typing, semantic units **bridge this gap**, serving simultaneously as structured models and as referential stand-ins (we will see this also applied in our approach for modelling causal relations below).

From a **FAIR Digital Object (FDO) perspective**, the compound unit can be serialized as a FDO, and its ID resource used consistently to refer to the real-world organism it describes. Typing it as both an '`organism description compound unit`' and an '`organism`'[OBI:0100026] enhances **semantic clarity**, **findability**, and **interoperability**, while supporting reasoning workflows that rely on class membership.

To make this dual function explicit, we recommend using a dedicated semantic relation such as '`represents`' or '`is about`'[IAO:0000136], which distinguishes the **representation-as-semantic-structure (intensional meaning)** from the r**epresentation-as-entity (extensional meaning)**. This creates a formalized **representation-referent link**, allowing semantic units to function as both interpretable knowledge artifacts and semantic proxies for the entities they describe.

# Modelling Causal Relations

## Correlation and Causation as Semantic Building Blocks

Causal inference in ecology often begins with observed correlations. These statistical associations suggest potential causal relationships, but do not themselves establish causality. To support flexible and scalable causal modelling, we introduce a **correlation schema metamodel** that accommodates both correlation and causation within a unified, logic-aware framework.

This metamodel defines a **directed binary relationship between two causal variables**, which may represent **qualities**, **dispositions**, or **processes**. With qualities, we refer to characteristics intrinsic to their bearer (material entity or process) that are typically observable and that the bearer possesses at a given time, such as the leaf nitrogen concentration of a particular plant at a specific time. Dispositions are inherent tendencies or capacities of their bearer that may only manifest under certain conditions, such as the drought tolerance of a particular plant. Finally, processes are dynamic occurrences or events that unfold over time, such as planting.

The model treats these variables as explicit entities, enabling them to serve as **observable carriers of potential causes and effects**. While causal graphs require directionality, the metamodel itself is **agnostic whether the relationship is correlative or causal**. Since causation implies correlation[2], but not vice versa, the set of correlative relationships includes the set of causal relationships, and correlation provides a general framework from which more specific causal refinements can be derived.

The semantic schema of the metamodel consists of three slots, i.e., Variable_1, Relationship, Variable_2, with the Relationship slot having `'correlated with'[RO:0002610]` as the default relationship:

[13]     S:`'variable_1'`,
         P:`'correlated with'[RO:0002610]`,
         O:`'variable_2'`

This structure defines a **correlation statement unit**, modelled as a semantic unit within the knowledge graph (Fig. 5).

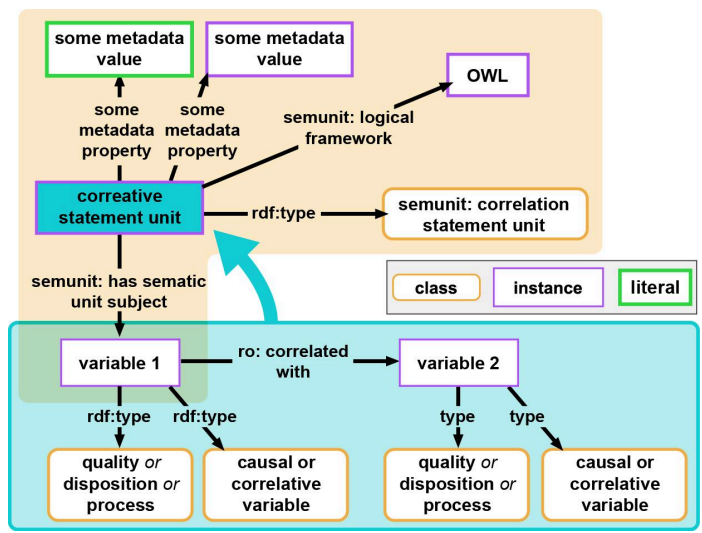

**Figure 5: Correlation schema metamodel as a correlation statement unit**. The correlation schema metamodel from [13] as a correlation statement unit.

For example, the correlation "*Competitive suppression of resident species on non-native species decreases invasion success*" is captured by the universal correlation statement:

---

[2] Two variables $X$ and $Y$ are correlated if: $P(Y \mid X) \neq P(Y)$. We thus understand correlation as any statistical dependence, and not only as a linear association between two variables. The statement "Causation implies correlation" refers to the ontological aspects of causation. Epistemologically, the statistical dependence (correlation) can be hidden by other variables (e.g., confounding can mask the effect, effects cancel out symmetrically, or measurement errors or aggregation hides dependence) so that observational data may fail to detect any correlation, the causal dependence still exists in the full causal model.

[14]     S: every-instance of `'competitive suppression of resident species on non-native species'`,
         P: `'negatively correlated with'[RO:0017004]`,
         O: some-instance of `'invasion success'`

Here, `'variable_1'` is a process and `'variable_2'` a disposition. The use of every-instance and some-instance ID resources reflects that the relationship is universal. The default `'correlated with'[RO:0002610]` relationship is replaced by a directed relationship.

To model a **causal interpretation** of this correlation, the relationship is replaced by `'negatively regulates characteristic'[RO:0019002]` (Fig. 6).

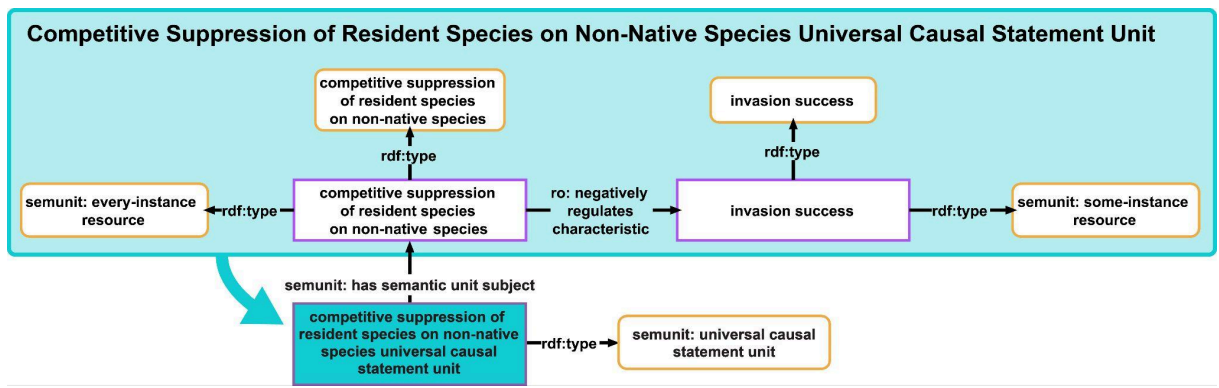

**Figure 6: Example of a universal causal statement unit**. The semantic unit modelling the causal hypothesis *"Competitive suppression of resident species on non-native species decreases invasion success"*. It represents the causal interpretation of the universal correlation statement from [14]. Notice the use of `'every-instance'` and `'some-instance'` ID resources with the two causal variables. Triples specifying the unit's metadata are not shown. The blue arrow indicates that the ID resource of the universal causal statement unit (blue box with purple border) represents the unit's semantic content (content-graph in the blue-bordered box).

The **universal causal statement unit** refines the prior observational or statistical correlation and asserts a mechanistic or explanatory claim. One can argue that the causal interpretation instantiates a semantic possibility space implied by the general correlative association of the two variables. The relation between the two statement units can be formally encoded using a `'causal interpretation of'` property. The causal interpretation does not invalidate the correlation but specializes it semantically.

Moreover, variables can themselves be modelled as semantic units or **dual typed** as both semantic units and an ontology class instance. In this way, `'variable_1'` in the above example may be modelled as a compound unit instantiating both a compound unit class and `'competitive suppression of resident species on non-native species'` (Fig. 7).

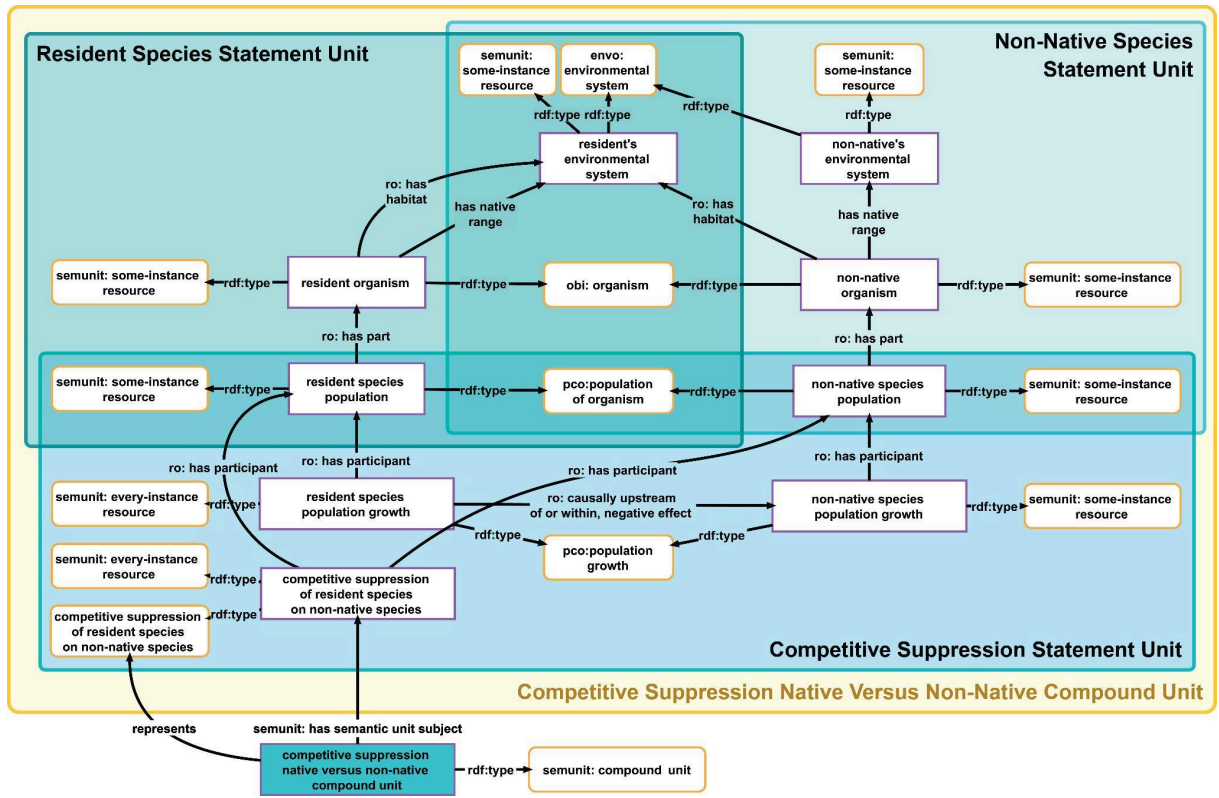

**Figure 7: Example of a causal variable with dual typing:** The causal variable from Fig. 6, modelled as a compound unit (yellow-bordered box) that comprises three statement units (blue-bordered boxes). To reduce the visual complexity, only the content-graphs of the compound unit and its statement units are displayed.

Causal hypotheses modelled as universal causal statements can be instantiated by empirical observations modelled as assertional causal statement units. When a specific assertion contains the triple shown in [15], it satisfies the universal causal hypothesis shown in Figure 6 and can be linked to it via the `'satisfies'` property. This connection provides both evidentiary grounding and a structure for inferential reasoning: the causal relationship provides an explanation for the observed decrease of the invasion success of the non-native species as an *effect* caused by the competitive suppression of resident species on non-native species.

[15]  S:`'competitive suppression of resident species on non-native species'`,
P:`'negatively regulates characteristic'`[RO:0019002],
O:`'invasion success'`

Universal causal statement units can also be qualified according to classical categories of causal strength, following Pearl's distinction between necessary, sufficient, and necessary-and-sufficient causes (21). A **necessary cause** is one without which the effect would not occur, but whose presence alone does not guarantee the effect (e.g., the presence of viable seeds may be necessary for invasion success, but not sufficient without suitable habitat). A **sufficient cause** is one whose presence guarantees the effect, though other causes may also lead to the same outcome (e.g., complete competitive release in the absence of native species might be sufficient for invasion success)). Some causes may be both **necessary-and-sufficient**, meaning the effect occurs if and only if the cause is present.

Within the Semantic Units Framework, these distinctions can be captured by defining corresponding subclasses of universal causal statement units (e.g., necessary universal causal statement unit). This semantic differentiation enhances interpretability, supports more precise causal modelling, and can inform both hypothesis refinement and evidence evaluation in ecological research.

## Composing Universal Causal Statement Units into Causal Networks

To form interpretable causal or correlative networks from individual statement units, these units must be logically composable. However, not all statement units can be combined arbitrarily. To preserve semantic coherence and avoid logical inconsistencies, we defined the following rules for combining two universal statement units, A and B, whether they express correlative or causal relations (cf., Fig. 8):

[16] **Combination Rules**

[16.1] The **target variable** (variable$_2$) of statement A must instantiate the same ontology class or semantic unit class as the **source variable** (variable$_1$) of statement B. This ensures the two statements share a common connecting variable and can thus be semantically chained.

[16.2] If the target variable of A uses a some-instance ID resource, then the source variable of B must use an every-instance ID resource. In such cases, and if rule [16.1] applies, it is safe to replace the every-instance ID resource in B with the some-instance ID resource from A when constructing the composite unit. This aligns with classical logical entailment: what applies to every instance of a class also applies to any specific instance.

As an example, consider two universal causal statement units:

[17] **Statement A** (cf. Fig. 6)

S: every-instance of `'competitive suppression of resident species on non-native species,`

P: `'negatively regulates characteristic'` [RO:0019002],

O: some-instance of `'invasion success'`

[18] **Statement B**

S: every-instance of `'niche differentiation between non-native and native species',`

P: `'causally upstream of or within, negative effect'` [RO:0004046],

O: some-instance of `'competitive suppression of resident species on non-native species'`

The two statements satisfy both Rule [16.1] and Rule [16.2], since the target of B matches the source of A by class identity, and the replacing the ID resource of the target of B with the source of A is logically valid. Therefore, these statements can be composed to form a larger causal path (Fig. 8), enabling semantic traversal through a causal chain.

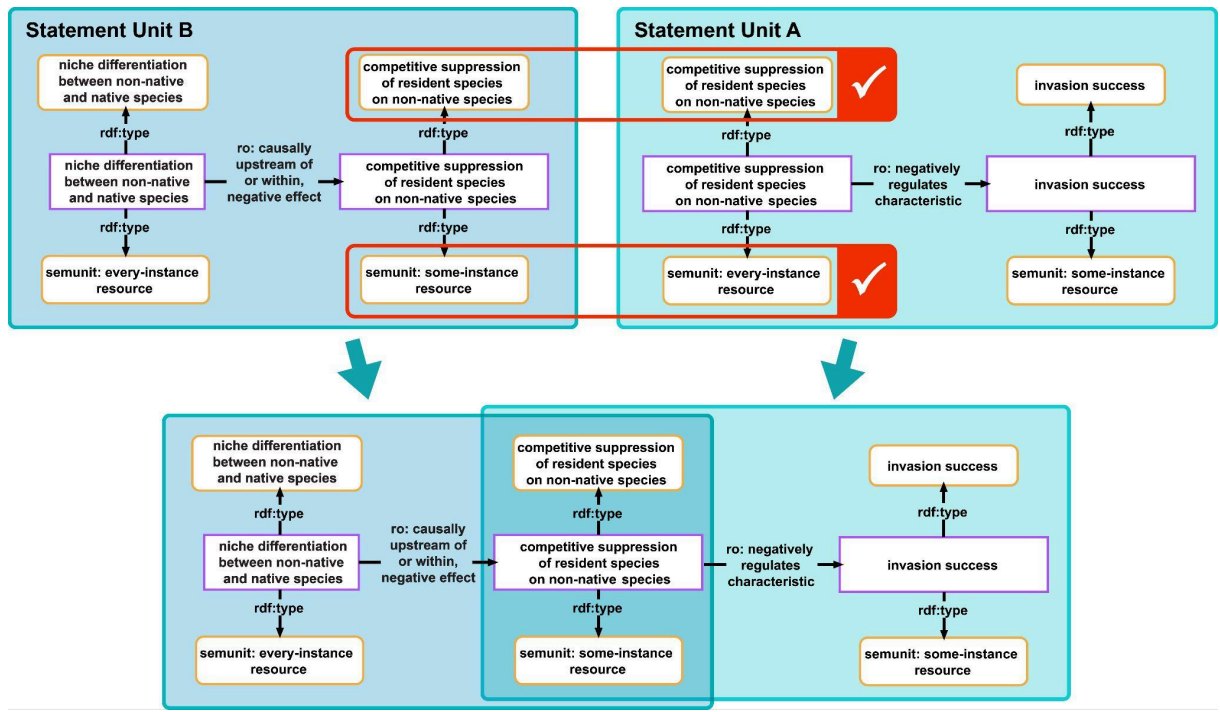

**Figure 8: Combining two universal causal statement units to form a causal chain. Top:** Two universal causal statement units that satisfy the rules [16] (highlighted in red). **Bottom:** The combination of the two statement units.

## Causal Maps: Navigating Causal Relationships in Ecological Knowledge

By combining universal causal statement units as described above, complex networks of causally or correlatively linked variables can be constructed. For example:

[19] **Statement C**

S: every-instance of `'non-native species fit to the habitat'`,

P:`'causally influences, positive effect'`,

O: some-instance of `'invasion success'`

[20] **Statement D**

S: every-instance of `'niche differentiation between non-native and native species'`,

P:`'negatively regulates characteristic'`[RO:0019002],

O: some-instance of `'non-native species fit to the habitat'`

These and the Statements A and B together form a directed causal network. Such a network can be formalized as a **compound unit** (Fig. 9) representing a higher-level semantic structure, i.e., a **causal map**.

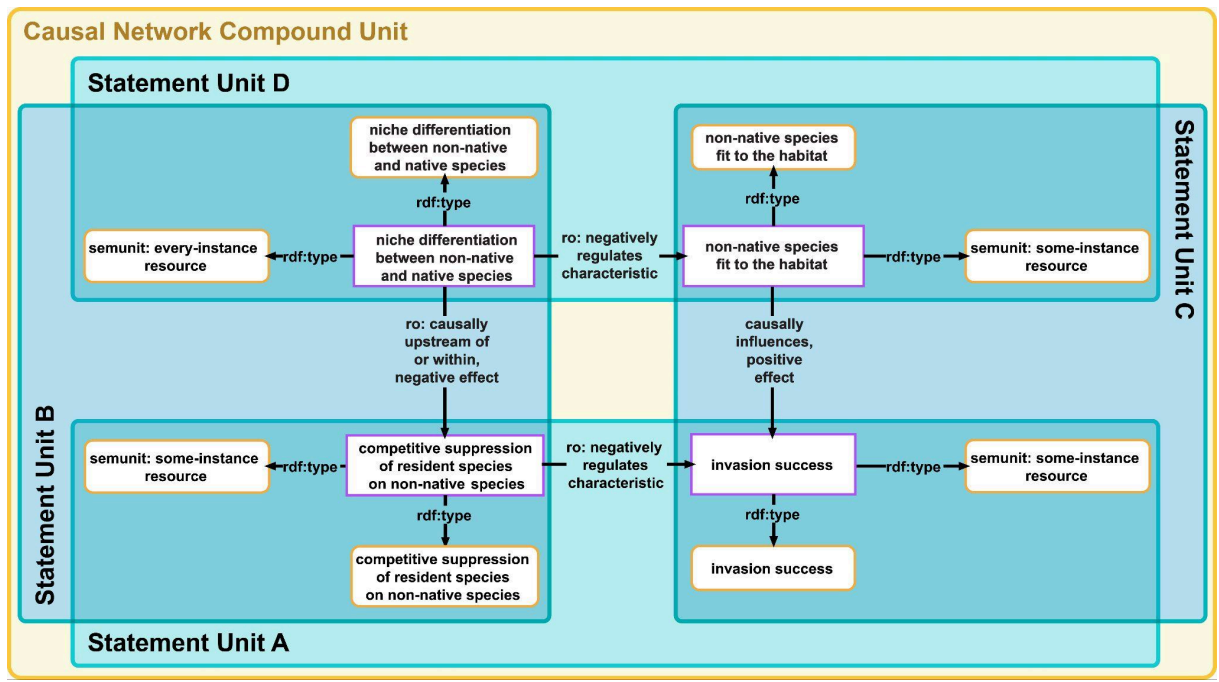

**Figure 9: A causal network consisting of the combination of four universal causal statement units**. A causal network compound unit (yellow-bordered box), consisting of four universal causal statement units (blue-bordered boxes), which have been combined following the rules from [16].

Following this procedure, all universal causal statement units in a knowledge graph can be combined, following the rules stated in [16], to form a corresponding compound unit that represents a causal network. The same can be done for the correlation statement units. The two resulting networks can be aligned via the `'causal interpretation of'` property to form a single map that functions like a **semantic grid**, analog to the spatial grid of map services such as [OpenStreetMap](#), to which additional information can be pinned at **specific locations on that map**. Each binary relationship (i.e., each correlative or causal statement unit) has, with its particular ID resource, a **unique address** that can be **located on the map**. The ID resources thus serve as addressable anchors to which additional information may be linked. By understanding each causal relationship as a **scientific claim**, such additional information may include:

- Empirical observations (assertional causal statement units) as evidence supporting or contradicting a causal claim, via the `'satisfies'` property.
- Scholarly references or DOIs that support or contradict a causal hypothesis.
- Experimental metadata and contextual qualifiers.
- Citation snippets or figures (where licensing permits).

By specifying variable types (categorical, ordinal, continuous), scales (interval, ratio), and value ranges (e.g., [0.0, 1.0]), as well as functions for the relationship between the variables (linear, quadratic, sinusoidal), the causal network can be interpreted within:

- **Directed Acyclic Graph (DAG)**, if the network contains **no cycles** (i.e., no variable can be an ancestor of itself),
- **Bayesian frameworks**, via conditional probabilities, and
- **Structural Equation Models (SEMs)**, via structural coefficients and endo-/exogenous distinction.

Causal maps and universal causal statement units can also be linked to quantitative models using Pearl's **structural causal models (SCMs)** (21). In an SCM, each variable is expressed as a mathematical function of its direct causes plus a stochastic term, defining how causal influences combine to determine outcomes. For instance, invasion success could be modelled as a logistic function of competitive suppression and habitat fit. Within the Semantic Units Framework, these structural equations can be represented as semantic units linked to the corresponding universal causal statement units, allowing the causal map to serve both as a human-interpretable knowledge model and as a scaffold for quantitative simulation, prediction, and counterfactual analysis.

This growing, navigable causal map with associated additional information, built on semantic units, supports hypothesis tracking, knowledge synthesis, and decision-making. It has the potential to become a valuable resource for researchers and practitioners to identify **causal explanations** for observed phenomena, and for ecosystem management to **predict possible effects** by identifying their potential causes. As each causal unit is a formal, addressable entity, it enables modular updates, contextual embedding, and integration with reasoning tools. The result is a dynamic, extensible causal landscape grounded in formal semantics and tailored to ecological research.

## Causal Junctions, Causal Perspectives, and Contextuality

**Causal Junctions**

In any causal network, two connected causal relationships form a **causal junction** involving three variables (i.e., $V_1$, $V_2$, and $V_3$). These junctions represent **structurally meaningful configurations** that underlie causal reasoning and explain inter-variable dependencies. They fall into three fundamental types (21):

- **Chain junctions** follow the pattern $V_1 \rightarrow V_2 \rightarrow V_3$, where $V_2$ **mediates** the causal influence from $V_1$ to $V_3$. In such a case, $V_2$ *screens off* $V_1$ from $V_3$, meaning that once $V_2$ is accounted for, knowing $V_1$ provides no additional predictive power about $V_3$. In the causal network illustrated in Figure 9, statement units D and C exemplify a chain junction.
- **Fork junctions** follow the pattern $V_1 \leftarrow V_2 \rightarrow V_3$, where $V_2$ is a **common cause** of both $V_1$ and $V_3$. This configuration may induce a statistical correlation between $V_1$ and $V_3$, even in the absence of any direct or indirect causal path between them. In Figure 9, statement unit B combined with the chain formed by D and C constitutes such a fork junction.
- **Collider junctions** follow the pattern $V_1 \rightarrow V_2 \leftarrow V_3$, where two independent causes converge on a shared effect $V_2$. Importantly, conditioning on the collider ($V_2$), e.g., through selection or stratification, can induce a spurious association between $V_1$ and $V_3$. In Figure 9, statement units A and C together form a collider junction.

These three junction types form the structural backbone of causal reasoning in DAGs, SEMs, and Bayesian networks. Junctions can be automatically identified and classified by adequate graph-traversal algorithms or reasoning procedures. Once identified, they can be **formally documented** in the knowledge graph as compound units of type **chain**, **fork**, or **collider junction unit**, which allows formal partitioning of the causal map into logically consistent and semantically interpretable subgraphs that are causally meaningful, forming modular components that support deeper interpretations and reuse.

**Mediation in the Semantic Units Framework**

In many ecological scenarios, the influence of one variable on another occurs not directly, but through an intermediate variable, a structure known as **mediation**. For example, habitat alteration might affect predator abundance, which in turn influences prey population size. Pearl's formalization of mediation distinguishes the **direct effect** of a candidate cause $V_C$ on an effect $V_E$ from its **indirect effect**, which operates through a **mediator** $V_M$ (22–24). In a causal diagram, this forms a chain junction $V_C \rightarrow V_M \rightarrow V_E$.

Pearl distinguishes several related quantities: The **total effect (TE)** of $V_C$ on $V_E$ includes both its **direct influence** and any **indirect effects** that operate through one or more mediators $V_M$. The **controlled direct effect (CDE)** isolates the effect of $V_C$ on $V_E$ while **holding $V_M$ fixed** at a specific value, reflecting hypothetical interventions. In contrast, the **natural direct effect (NDE)** is the effect of $V_C$ on $V_E$ when $V_M$ is allowed to vary naturally as it would if $V_C$ were set to a reference value. The **natural indirect effect (NIE)** is the proportion of the effect of $V_C$ on $V_E$ that operates entirely through $V_M$.

Under standard identification assumptions (22,24) that (i) **no unmeasured confounding** of the $V_C \rightarrow V_E$ relationship, conditional on observed covariates $Z$, exists, (ii) **no unmeasured confounding** of the $V_M \rightarrow V_E$ relationship, conditional on $V_C$ and $Z$, exists, and (iii) **no mediator-outcome confounders** affected by $V_C$ exist. The natural direct effect is given by Pearl's mediation formula:

[21]  **NDE** $= \sum_m [E(V_E \mid V_C = 1, V_M = m) - E(V_E \mid V_C = 0, V_M = m)] \, P(V_M = m \mid V_C = 0)$

The **natural indirect effect** is:

[22]  **NIE** $= \sum_m E(V_E \mid V_C = 0, V_M = m) [P(V_M = m \mid V_C = 1) - P(V_M = m \mid V_C = 0)]$

These satisfy:

[23]  **TE = NDE + NIE**

Within the Semantic Units Framework, mediation structures are explicitly represented as **chain junction units** linking $V_C$, $V_M$, and $V_E$. Direct and indirect effects are annotated on these junctions, specifying whether they represent total, controlled, or natural effects. Identification assumptions can be encoded as metadata linked to the causal perspective unit (see next section), for instance as lists of measured confounders or as observability of a mediator. Moreover, the formulas above can be stored as computational annotations, enabling automated reasoning engines to estimate NDE and NIE when the required data are present.

Because each causal relationship, mediator, and assumption is represented as a machine-interpretable semantic unit, the framework can integrate mediation analysis with front-door and back-door criteria, and even with do-calculus-based transformations (see next two sections), to estimate mediated effects in settings where direct paths are confounded or unobservable. By capturing mediation, direct, and indirect effects semantically, the framework enables researchers to not only determine whether a causal relationship exists, but also **how it operates**, opening the door for mechanistic insights and transparent ecological explanation.

**Causal Perspectives, Contextuality, and Deconfounding**

Researchers often focus on a **specific causal question**, such as "*How does variable $V_C$ (candidate cause) influence $V_E$ (candidate effect)?*" A **causal perspective** is a subnetwork centered around such a **focus-pair of variables $V_C$ and $V_E$** and includes all paths connecting $V_C$ to $V_E$ on the causal map as a collection of corresponding universal causal statement units, as well as all optional metadata and annotations associated with these paths, such as supporting or contradicting evidence and scholarly publications mentioning the causal hypotheses relevant to the paths and the focus-pair, as well as any annotations regarding chains, forks, and colliders. This collection forms a compound unit of type **causal perspective unit**, which can be formally represented in the knowledge graph as a queryable, FAIR, CLEAR, and reusable semantic unit. A causal perspective unit can thus act as a **semantic filter**, reducing the map's complexity to the information relevant to a particular research question or intervention.

Causal perspectives can be further refined by contextual dimensions, considering other criteria, such as taxonomic filters (e.g., only involving amphibians), ecosystem types (e.g., freshwater wetlands), or evidence source (e.g., controlled experiments). Such **context-sensitive perspectives** are serialized as formal compound units of the type **contextual causal perspective unit**, and become addressable and queryable entities within the knowledge graph. They can be cited, reused, extended, annotated, or compared, making them interoperable FAIR knowledge objects.

To estimate causal effects, correlations such as ($P(V_E | V_C)$) are not sufficient. We require the interventional probability $P(V_E | do(V_C))$, where the **do-operator** signifies that $V_C$ is set experimentally, removing all incoming causal influences (21).

Pearl's criteria (21) define when **observational data** suffice for causal inference:

1. **Back-Door Criterion:**

    A set of variables $Z$ blocks all **back-door paths** (paths entering $V_C$ through arrowhead) and conditioning on $Z$ allows estimating the causal effect:

    $P(V_E | do(V_C)) = \sum_Z P(V_E | V_C, Z) \, P(Z).$

    In the Semantic Units Framework, algorithms can identify all back-door paths for a given causal perspective, compute minimal sufficient adjustment sets, and store this as a **back-door causal perspective unit**.

1. **Front-Door Criterion:**

    A mediator $V_M$ transmits all causal influence from $V_C$ to $V_E$. Conditions for valid front-door adjustment are that $V_C$ influences $V_E$ only through $V_M$, there are **no unblocked back-door paths from $V_C$ to $V_M$**, and all **back-door paths from $V_M$ to $V_E$ are blocked** by conditioning on $V_C$. If these conditions are met, the causal effect can be computed using:

    $P(V_E | do(V_C)) = \sum_{VM} P(V_M | V_C) \sum_{V'C} P(V_E | V_M, V'_C) \, P(V'_C).$

    Within the Semantic Units Framework, mediators are identified from the causal map as chains of universal causal statement units, and their status as front-door candidates can be semantically annotated and reasoned over. They are represented in the knowledge graph as **front-door causal perspective units**.

3. **Instrumental Variables:**
   Some causal effects can only be identified through an **instrumental variable $Z$**, which influences $V_C$ (relevance), but has no direct path to $V_E$ except through $V_C$ (exclusion), and is independent of unobserved confounders. This is represented in the knowledge graph as **instrumental-variable causal perspective unit**.

**Do-Calculus and d-Separation: Generalizing Causal Inference**

Back-door, front-door, and instrumental variables rules are **special cases of Pearl's do-calculus**, which allows more general reasoning over **interventions in DAGs**. (21).

d-Separation is a graphical criterion that determines whether two variables $V_C$ and $V_E$ are conditionally independent given a set of variables $Z$. Two variables are **d-separated by $Z$**, if every path between them is **blocked** when conditioning on $Z$, following three basic path types:

1. Chains ($V_C \rightarrow V_M \rightarrow V_E$) are blocked if $V_M$ is in $Z$.
2. Forks ($V_C \leftarrow V_M \rightarrow V_E$) are blocked if the common cause $V_M$ is in $Z$.
3. Colliders ($V_C \rightarrow V_M \leftarrow V_E$) are blocked by default but become open if the collider $V_M$ or any of its descendants is in $Z$, which can induce spurious correlation.

The do-calculus is a general framework of three rules that transform expressions with interventions (e.g., $P(V_E \mid do(V_C))$) into observational expressions, guided by d-separation in modified graphs. The three do-calculus rules are:

- **Rule 1: Insertion/Deletion of Observations**
  Observing a variable $W$ that does not open any blocked path (per d-separation) does not affect the causal expression: $P(V_E \mid do(V_C), Z, W) = P(V_E \mid do(V_C), Z)$
- **Rule 2: Intervention/Observation Exchange**
  If $V_C$ and $V_E$ are d-separated given $Z$ after removing incoming edges to $V_C$, then: $P(V_E \mid do(V_C), Z) = P(V_E \mid V_C, Z)$
- **Rule 3: Insertion/Deletion of Interventions**
  If intervening on a $Z$ does not affect $V_E$ given the rest of the graph (by d-separation), that intervention can be ignored: $P(V_E \mid do(V_C)) = P(V_E)$

The Semantic Units Framework is well suited for **semi-automated** application of the do-calculus because the causal maps represent semantic DAGs built from universal causal statement units, with causal junctions being represented as explicit semantic unit types, and the (contextual) causal perspectives providing focused subgraphs for a given $V_C \rightarrow V_E$ relationship. Moreover, back-door, front-door, and instrumental variable causal perspectives can store adjustment sets derived via do-calculus.

With these elements, the Semantic Units Framework, supported by adequate algorithms, can identify confounders, mediators, and colliders for a focal causal question, evaluate d-separation to determine valid adjustment sets, detect back-door and front-door paths automatically, apply the do-calculus rules to generate simplified interventional expressions, and document derived causal perspectives as compound units linked to their sources via a property such as `'derived by do-calculus from'`.

This transforms ecological causal maps into computationally actionable DAGs, bridging semantic knowledge representation with formal causal inference. It enables ecologists to move from observational correlations to justified interventional claims, supporting knowledge synthesis, decision-making, and reproducible AI-ready workflows.

## Counterfactual and Potential Outcome Reasoning in the Semantic Units Framework

Causal inference in ecology often requires answering counterfactual questions, such as "*Would the invasion success of a non-native species have been lower if native species density had been higher?*" or "*Would wetland restoration have increased amphibian richness if invasive plants had been removed?*" These questions move beyond observational or interventional statements to **counterfactual reasoning**, which involves reasoning about **alternative realities** relative to the observed one.

In Pearl's framework (21), counterfactual reasoning is formalized using **potential outcomes**, where $Y_x$ denotes the value of an outcome $Y$ had a variable $X$ been set to a value $x$ via intervention. While back-door and front-door criteria identify conditions under which causal effects can be estimated from observational data, **counterfactuals** allow us to reason about individual or scenario-specific "*what-if*" statements.

Within the Semantic Units Framework, counterfactuals can be modelled as **counterfactual statement units**. As counterfactuals represent logically possible but non-actual scenarios derived from interventions applied to the causal model, they represent a subtype of contingent statement units. They describe possible worlds consistent with the model rather than actual observed events. Counterfactual statement units are linked to their corresponding universal causal statement units via the `'satisfies'` property, enriched with **counterfactual context annotations** and combined to a **potential outcome compound unit** consisting of:

1. **Observed outcome statement unit** and thus the assertional causal statement unit that documents the factual scenario (e.g., observed invasion success for a given plot).
2. **Counterfactual statement unit**, representing the hypothetical outcome under an alternative intervention (e.g., predicted invasion success if competitive suppression were higher).
3. **Universal causal statement unit** that underlies the hypothetical reasoning.

This approach allows the causal map to encode both factual and hypothetical knowledge. Potential outcome compound units can be further annotated with intervention descriptions (e.g., *do*(variable = value)), model or method used to derive the counterfactual (e.g., Bayesian network inference, simulation model, or expert assessment), and evidence level or uncertainty.

By integrating potential outcomes into causal perspectives, the Semantic Units Framework supports decision-support workflows, as ecologists can query "*Which variables, if changed, would most likely improve habitat restoration success?*" and managers can explore "*Which combination of interventions is predicted to prevent invasive species dominance under current conditions?*".

Formally representing counterfactuals as semantic units provides a **bridge between observed data**, **causal maps**, and **decision-relevant scenario reasoning**, extending the framework from knowledge synthesis to actionable ecological insights.

# Modelling Unobservable Qualities: Diagnostics, Methods, and Semantic Representations

Many critical ecological variables and processes, such as phylogenetic relatedness, niche overlap, dispersal potential, or resilience, are **not directly observable**. Instead, they are inferred through models, diagnostics, or proxies based on empirical data and assumptions. These inferences are often context-specific and depend heavily on the methods applied, leading to multiple, sometimes competing, representations of the same underlying entity. Here, we explore how the Semantic Units Framework can account for such unobservable yet scientifically important constructs.

## Unobservables and Method-Dependent Variation

Consider the example of *phylogenetic relatedness between non-native and native species*, a relational quality that plays an essential role in invasion ecology. Its quantification is not direct but depends on analytical methods and modelling assumptions. Two possibilities are:

- **Method A**: Mean phylogenetic distance between a non-native species and each native species in the community, weighted by the abundance of those natives (i.e. abundance-weighted mean phylogenetic distance; awMPD).
- **Method B**: Phylogenetic distance between a non-native species and its nearest taxonomic relative in the native community (nearest taxon distance, NTD).

Both methods aim to represent the same underlying quality, but they rely on different data aggregation techniques, assumptions about ecological relevance, and statistical models. Consequently, the resulting values are not interchangeable, and their interpretation is only meaningful within the context of their respective methodologies.

## Two Modelling Approaches

Within the Semantic Units Framework, we propose two complementary strategies to represent such method-dependent constructs:

**Reflective Indicator Approach: Contextualizing the Measurement**
In this strategy, the unobservable quality (e.g., *phylogenetic relatedness*) is defined as a class in a corresponding ontology, and each measurement or estimate of corresponding instances of this ontology class are modelled as an assertional statement unit that documents in its metadata via a `'measured/estimated applying method'` property which method has been used to obtain it. Alternatively, the entire measurement process can be modelled in a set of corresponding statement units, documenting which methods have been applied and which data processing steps have been conducted to obtain the measurement or estimate.

This reflective indicator strategy preserves the ontological identity of the underlying quality and clearly specifies how each observed value was derived, supporting traceability and reproducibility. It treats the measurement as the outcome of a **direct observation of the underlying quality** (i.e., reflective indicator).

**Formative Indicator Approach: Method-Defined Subclasses**
Alternatively, one may treat each methodologically defined version of a construct as a distinct subclass of the general quality class. This recognizes that, in practice, the method defines what is being measured (i.e., formative indicator). In this case, the general quality `'phylogenetic relatedness'` would have subclasses such as `'phylogenetic relatedness measured as abundance-weighted mean phylogenetic distance'` (Method A) and `'phylogenetic relatedness measured as nearest taxon distance'` (Method B).

This strategy is especially useful when the methods are widely standardized and considered epistemically distinct, when different methods are not commensurable (e.g., apply to different data structures), and when downstream reasoning or querying must treat them differently. The formative indicator approach enhances semantic precision, but risks fragmenting the conceptual space if not managed carefully.

The decision between the reflective and the formative indicator strategy should be informed by epistemic stance, pragmatics, and community standards. Regarding **epistemic stance**, the question is do researchers consider the underlying quality to be a single conceptual target with methodological contextualized observations (reflective approach) or do they treat each methodological variant as a distinct operational construct (formative approach)? Regarding **pragmatics**, the question is whether the domain benefits from generalized queries across methods, or does it require strict method-based distinctions? The question relating to **community standards** is whether certain methods are recognized as authoritative or canonical?

In practice, both strategies can coexist. The reflective approach may dominate in early-stage or exploratory research, while the formative approach may emerge in domains where measurement standards become institutionalized.

Regarding ontology development, the modelling challenge underscores the importance of distinguishing between **ontological realism** and **epistemic contextualism** in semantic modelling. Ontologies need to accommodate both the stable identity of the referent (e.g., a relational quality like phylogenetic relatedness) and the method-dependent ways in which this referent is constructed or estimated.

The Semantic Units Framework, with its support for modularization of units of information, dual typing, and contextual metadata, provides the necessary flexibility. By using compound units to nest method-bound observations and linking them to general qualities via properties like `'measured/estimated applying method'`, we can maintain a coherent representation space that respects both the semantics of the data and the practices of ecological science.

**Example of Linking a Method-Defined Causal Variable and a Corresponding Metric**
*Invasion success* serves as a useful example of an unobservable ecological quality that can function as a causal variable in a hypothesis (Fig. 10, Top). Since invasion success is not directly measurable, it must be operationalized using methods that define how it is indirectly assessed. One such method might involve quantifying the above-ground dry biomass of the invasive population within a given area.

Following the **formative indicator approach**, we model `'invasion success measured as above-ground dry biomass of invasive population'` as a **method-defined subclass** of `'invasion success'`. This subclass encapsulates a specific operationalization that ties the concept variable to a particular empirical strategy.

This subclass can then be referenced in assertional causal statement units when modelling concrete hypotheses in a given ecological setting (Fig. 10, Middle). These assertional units, in turn, can be linked to the corresponding measurement results (e.g., a numeric biomass value in proportion to a given area) documented in a related assertional measurement statement unit (Fig. 10, Bottom).

The assertional statement unit may then be linked to its universal causal statement unit via the property `'satisfies'`, and both assertional and universal causal statement units may reference the corresponding assertional measurement statement unit via the property `'has associated measurement'` (Fig. 10).

This pattern ensures semantic clarity, allows for different operationalizations of the same conceptual variable, and preserves traceability from abstract causal structure to empirical data.

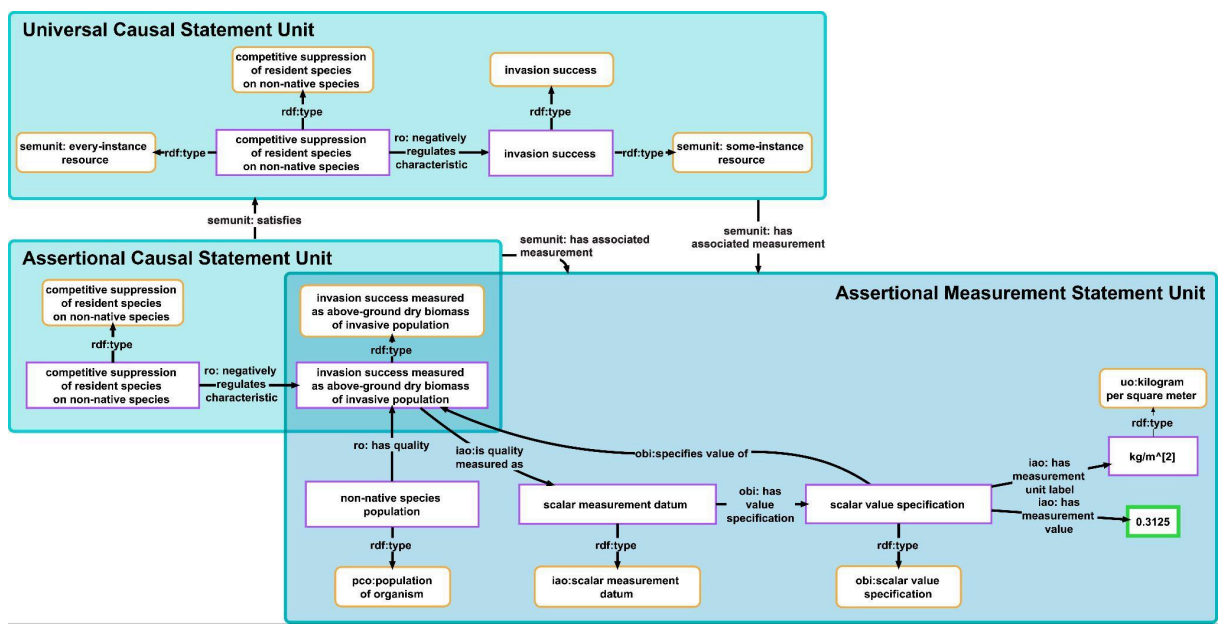

**Figure 10: Example of linking a method-defined causal variable and a corresponding metric. Top**: A universal causal statement unit representing a causal hypothesis. **Middle**: An assertional causal statement unit that corresponds with the causal hypothesis from above, indicated by linking the two via the `'satisfies'` property. **Bottom**: An assertional measurement statement unit with the metric for the effect variable of the assertional causal statement unit. In both the assertional causal statement and the corresponding measurement statement, the invasion success of the non-native species population is modelled using the `'invasion success measured as above-ground dry biomass of invasive population'` class, instead of its parent class `'invasion success'`, because it was assessed by measuring the dry biomass of the non-native species population for a given area.

## Semantic Units and the CLEAR Principle

A semantic unit is a self-contained, semantically meaningful building block of a knowledge graph. Each unit represents a single, logically coherent proposition, or a closely connected group of propositions, that can be independently identified, interpreted, reasoned over, queried, and reused. To maximize their utility for both humans and machines, semantic units are designed to align with the CLEAR Principle.

The **CLEAR Principle** (14) articulates a set of five interlocking goals for how to model and document data and knowledge: data and knowledge should be **C**ognitively interoperable, semantically **L**inked and contextualized, easily **E**xplorable, intuitively **A**ccessible, and **R**eadable and interpretable to humans. Within the Semantic Units Framework, these goals are operationalized through three design criteria (14):

**Criterion 1: Semantic Granularity and Identity**
Each statement unit represents exactly one proposition and is assigned a globally unique and persistent identifier. This guarantees referential integrity and allows the unit to be directly addressed, referenced, and queried. This design supports semantic precision and conforms to FDO principles.

**Criterion 2: Nested Composability**
Semantic units can be hierarchically composed into more complex structures. For example, multiple statement units can be grouped into a compound unit representing a causal network, a multi-step observation process, or a research hypothesis. Each compound unit is itself assigned a unique identifier and instantiates a corresponding compound unit class. This enables modular modelling and supports serialization as **nested Nanopublications** or **nested RO-Crates**, aligning with FAIR and machine-actionable knowledge infrastructures.

**Criterion 3: Human-Centric Presentation**
Machine-readable and -interpretable structures are decoupled from their human-facing representations. Interfaces built on semantic units can render those elements to a user's task or cognitive context (e.g., using dynamic labels and dynamic graphs, see Fig. 2D,E). This increases interpretability and supports exploratory workflows by offering context-sensitive navigation and intuitive visualizations. In this way, semantic units support **cognitive interoperability** and **context-dependent exploration** (see (14)) of data and knowledge, enhancing human comprehension without sacrificing formal semantic structure.

Together, these criteria ensure that each semantic unit is both semantically interoperable, via ontology-based modelling, and cognitively interoperable, through human-centered design. The Semantic Units Framework thus supports the dual imperative of machine-actionability and human-intelligibility, and thereby provides semantic models not only for **FAIR** but also **CLEAR data and knowledge representations**.

Moreover, the principles of granularity, composability, and transparency are consistent with the prior section's treatment of causal modelling, unobservable qualities, and representation-referent identity, reinforcing the framework's coherence and extensibility.

# Discussion [WIP]

Given that this seems a system for self-referencing knowledge: How do all these relate to Gödel's incompleteness theorem?

*The semantic content of each semantic unit models some aspect of (ecological) reality. Like all models, it is necessarily incomplete: it abstracts from its referent system to emphasize only those features relevant to its intended purpose and context of use. This is not a limitation of the framework but an inherent property of all modelling. The Semantic Units Framework explicitly accommodates this incompleteness by linking each unit to its evidence, methods, and scope, enabling the knowledge graph to evolve as new observations and interpretations emerge.*

*In formal terms, any logical representation, whether in Description Logics (OWL) or*

*First-Order Logic, is inherently limited in the scope of truths it can capture and infer. In more expressive settings, Gödel's incompleteness theorem implies that there will always be true statements about the world that cannot be proven within the system. In less expressive but decidable logics, like OWL DL (which is a particular OWL language from the OWL language family—there are different flavors of OWL, depending on which type of Description Logic it is based on), expressivity is deliberately restricted to ensure reasoning can be computed, but at the cost of being unable to represent certain statements at all. Beyond these formal limits, there is the inherent modelling incompleteness (see above).*

Beyond ecology: how can this method be used in other fields of knowledge? Challenges & possibilities of interoperability and linking knowledge graphs.

*Although demonstrated here in an ecological context, the Semantic Units Framework is domain-agnostic and applicable across diverse fields such as medicine, social sciences, engineering, and climate change. Its modular design, representing assertional, contingent, prototypical, and universal statements as FAIR Digital Objects linked to ontologies, enables knowledge graphs from different domains to be aligned, connected, and reasoned over in a shared causal framework. This supports interoperability, cross-domain evidence synthesis, and integrated causal reasoning, while accommodating domain-specific terminologies, granularity levels, and evidence standards.*
*This, however, requires that the domain-ontologies referenced in semantic units across different domains are semantically interoperable with each other. They must thus align in their core concepts and relations and avoid contradictory ontological commitments (i.e., be based on the same foundational ontology, such as OBO ontologies are all based on the Basic Formal Ontology), and the semantic units classes that model a specific type of semantic content must use the same semantic schemata for modelling that content and must be applicable across different domains (if that type of content is relevant in multiple domains, such as for instance measurements).*

How to integrate CARE principles?

*The FAIR and CLEAR Principles address the technical, semantic, and cognitive aspects of ecological data and knowledge management, but they do not explicitly cover the social, ethical, and governance dimensions. The CARE Principles* (25) *for Indigenous Data Governance (Collective Benefit, Authority to Control, Responsibility, Ethics) complement FAIR and CLEAR by ensuring that data use respects the rights, interests, and values of Indigenous Peoples and local communities. Within the Semantic Units Framework, CARE can be operationalized by embedding governance and ethical constraints directly into the knowledge graph: statement and compound units can carry machine-readable CARE annotations specifying intended benefits, authority to control, responsibilities, and ethical requirements. Ontologies can be extended with CARE-specific properties, and reasoning rules can enforce compliance with declared governance terms. In this way, the framework supports not only interoperability and cognitive accessibility, but also socially just and ethically grounded ecological knowledge management.*

How to use the Framework for causal inference?

*Gather all known potential variables for a given effect, e.g., by looking at the bearer of the effect and all entities that bearer can potentially interact with. Then identify (or create) the semantic model for these potential variables and specify them within the knowledge graph. The application then identifies all possible causal/correlative relationships and draws an initial causal map. By plausibility and background knowledge, domain experts identify relevant from irrelevant causal relationships and their directionality. From here, empirical data and case-study information can be associated with the causal map, and do-calculus can be applied, identifying different types of junctions, etc.*

How can the application and knowledge graph support transportability of a causal map based on a specific use case to a more general applicability?
*Support of aligning and mapping components (variables) of several particular use cases to support their generalization. Providing ways to adapt a particular causal map to a new use case (identifying differences between them and adjusting the causal map accordingly), to then gather information about this use case and do some graph-adjustments etc., adding, e.g., new relationships (new variables) and deleting those not relevant in the given use case.*

> *Also, the do-calculus can help in the transportation process, by performing a valid sequence of do-operations to transform the target quantity into another expression in which the case-distinguishing variable is free of do-operators (see Bareinboim's algorithms).*

> *Ideally, a practitioner describes what they have done ("I have done X = x, and the outcome was Y = y" while specifying a target outcome y', the system answers that by doing X = x', the outcome would have been Y = y'.*

Semantic Units Framework, their knowledge graphs, and AI systems: how do they mutually benefit from each other?
*The Semantic Units Framework offers a structured, semantically explicit representation of ecological knowledge that is directly actionable by AI systems. Conversely, AI can both consume and enhance knowledge graphs based on this framework, creating a feedback-loop that strengthens both the causal reasoning capabilities of AI and the comprehensiveness and accuracy of the knowledge graph itself.*
**Benefits for AI systems**: *With API-level access to a knowledge graph enriched with causal maps, back-door/front-door path annotations, do-calculus-ready DAG structures, and linked empirical evidence (i.e., a knowledge graph based on the Semantic Units Framework), AI systems can:*
- *start from a validated causal framework instead of inferring causal structures entirely from raw data, improving reliability and reducing false positives;*
- *query causal pathways, identify adjustment sets, and compute causal estimates by applying formal rules (back-door, front-door, do-calculus) over well-defined graph structures;*
- *evaluate evidence provenance, methodological context, and uncertainty directly linked to causal relationships, enabling evidence-aware decision-making;*
- *provide transparent, human-interpretable explanations of causal reasoning via the framework's perspective units, supporting trust and interpretability.*
**Benefits for the Semantic Units Framework**: *AI systems, in turn, can enhance the framework's knowledge graphs by:*
- *suggesting candidate causal relationships from new datasets or literature mining, which can be reviewed and validated by domain experts before integration;*

- *automatically populating semantic units with new empirical evidence, metrics, and methods, increasing coverage and timeliness;*
- *detecting inconsistencies, redundancies, or logical gaps in the causal maps via reasoning over large, cross-domain datasets;*
- *supporting adaptive granularity, by proposing context-specific simplifications or expansions of causal maps for targeted research or management questions.*

*The relationship between knowledge graphs based on the framework and AI system can form a virtuous cycle, with the framework providing the formal, high-quality causal backbone that AI needs for explaining reasoning, while AI expands, refines, and keeps the knowledge graph up to date. Over time, this enables more robust, evidence-based ecological models, accelerated discovery of causal mechanisms in complex socio-ecological systems, and seamless integration of ecological knowledge with other domains such as climate science, epidemiology, or economics.*

*By coupling the Semantic Units Framework with AI systems in this way, ecological research and decision support can move from descriptive analytics to predictive and prescriptive reasoning, with each side continuously enhancing the other.*

The knowledge graph should also allow for specifying alternative causal maps for the same underlying referent system - like different causal scenarios. But these should be specified as alternative causal maps and should specify the different assumptions that underlie them.

# Future Directions [WIP]

- Causality and complexity: the role of granularity modelled using formal semantics (granularity trees and granularity perspectives) for questions of generalizing and abstracting causal relationships, zooming in and out, context-dependence, and rules for graph operations on causal networks (complementing or in addition to Pearl's *do*-calculus)
- Combine the algorithms mentioned in the manuscript for causal reasoning and interventions (*do*-calculus) and the knowledge graph into an integrated application that can be used by both ecology researchers and ecology practitioners -> EcoWeaver

# Conclusion  [WIP]

# Term Definitions

***above-ground dry biomass***
**Definition:** "A mass of a material of biological origin that was collected from above-ground from a defined area and dried for 3 days at 65 °C."
**subclass of:** `'mass' [PATO:0000125]`

*competitive suppression*

**Definition**: "An ecosystem process in which the activities of individuals that are part of a population of organisms constrains or reduces the size of another population of organisms." ([dbXref](#))

**subclass of:** `'ecosystem process'` [ENVO:01001795]

*interspecific competition*

**Definition**: "A competitive suppression in which the activities of a population of one species constrains or reduces the size of the population of another species." ([dbXref](#))

**subclass of:** `'competitive suppression'`

*invasion success*

**Definition**: "An ecosystem process in which a non-native species establishes a self-sustaining population in an area where it has not historically maintained a self-sustaining population."

**subclass of:** `'ecosystem process'` [ENVO:01001795]

*invasion success measured as above-ground dry biomass of invasive population*

**Definition**: "An invasion success that is understood to be quantifiable in reference to the above-ground dry biomass of the invasive population."

**Axiom:** equivalent class: `'invasion success measured as above-ground dry biomass of invasive population' ≡ 'invasion success' AND 'has associated metric' SOME 'above-ground dry biomass'`

**subclass of:** `'invasion success'`

*niche differentiation process*

**Definition**: "An ecosystem process by which competing species use the environment differently in a way that helps them coexist." ([dbXref](#))

**subclass of:** `'ecosystem process'` [ENVO:01001795]

--------------------------------------------------------------------------------

*Operational Organism Unit (OOU)*

**Definition**: "An Operational Organism Unit (OOU) is a synthetic, statistically-constructed organism entity that represents the aggregated biological traits of the members of a real, defined population of organisms belonging to the same species, locality, or ecological context.

The OOU does not correspond to an actual physical organism, but serves as a semantic proxy or typified abstraction to which population-level trait summaries (e.g., mean body mass, average metabolic rate, median lifespan) can be associated in a structured, ontology-based data model."

***Operational Specimen Unit (OSU)***

**Definition**: "An Operational Specimen Unit (OSU) is a synthetic, statistically-constructed organism entity that represents the aggregated biological traits of the organisms contained in a real, defined specimen or material sample, typically comprising multiple individuals.

The OSU does not correspond to any particular physical organism, but serves as a semantic proxy to which pooled or average trait data (e.g., mean length, isotopic composition, biomass) derived from the specimen can be associated in a structured, ontology-based data model."

# Author Contributions

**Conceptualization**: L.V. (Semantic Units Framework and its adaption to ecology and the modelling of causal relations and causal networks as semantic units); J.I.B. (examples from ecology)
**Methodology**: L.V. (semantic units approach, semantic models and schemata)
**Writing—Original Draft**: L.V.
**Writing—Review & Editing**: All authors
**Visualization**: L.V.